\documentclass[accepted]{uai2023} % for initial submission
\usepackage[american]{babel}
\usepackage{natbib} % has a nice set of citation styles and commands
\usepackage{mathtools} % amsmath with fixes and additions
\usepackage{booktabs} % commands to create good-looking tables
\usepackage{tikz} % nice language for creating drawings and diagrams
\usepackage{comment}
\usepackage{amsmath}
\usepackage{amssymb}
\title{
Enhancing Treatment Effect Estimation: A Model Robust Approach Integrating Randomized Experiments and External Controls using the Double Penalty Integration Estimator
}

\author[1]{\href{mailto:<ycheng26@ncsu.edu>?Subject=Your UAI 2023 paper}{Yuwen Cheng}{}}
\author[2]{\href{mailto:<liliwu@microsoft.com>?Subject=Your UAI 2023 paper}{Lili Wu}{}}
\author[3]{\href{mailto:<syang24@ncsu.edu>?Subject=Your UAI 2023 paper}{Shu Yang}{}}

\affil[1]{%
    Statistics Dept.\\
    North Carolina State University
}
\affil[2]{%
    Microsoft Research NYC
}
\affil[3]{%
    Statistics Dept.\\
    North Carolina State University
  }

\newtheorem{theorem}{Theorem}
\newtheorem{assumption}{Assumption}

\newcommand{\be}{\begin{equation}}
\newcommand{\en}{\end{equation}}
\newcommand{\bea}{\begin{eqnarray}}
\newcommand{\ena}{\end{eqnarray}}
\newcommand{\ba}{\begin{array}}

\newcommand{\ea}{\end{array}}

\newcommand{\indep}{\perp \!\!\! \perp}

\newcommand{\T}{\mathrm{\scriptscriptstyle T}}

\newcommand{\E}{ {\mathbb{E}} } 
 
\newcommand{\V}{ {\mathbb{V}}} 
\newcommand{\bone}{\mathbf{1}}

\begin{document}
\maketitle
\begin{abstract}
Randomized experiments (REs) are the cornerstone for treatment effect evaluation. However, due to practical considerations, REs may encounter difficulty recruiting sufficient patients. External controls (ECs) can supplement REs to boost estimation efficiency. Yet, there may be incomparability between ECs and concurrent controls (CCs), resulting in misleading treatment effect evaluation. We introduce a novel bias function to measure the difference in the outcome mean functions between ECs and CCs. We show that the ANCOVA model augmented by the bias function for ECs renders a consistent estimator of the average treatment effect, regardless of whether or not the ANCOVA model is correct.  To accommodate possibly different structures of the ANCOVA model and the bias function, we propose a double penalty integration estimator (DPIE) with different penalization terms for the two functions. With an appropriate choice of penalty parameters, our DPIE ensures consistency, oracle property, and asymptotic normality even in the presence of model misspecification. DPIE is more efficient than the estimator derived from REs alone, validated through theoretical and experimental results.
\end{abstract}

\section{Introduction}\label{sec:intro}
Randomized experiments (REs), which allow researchers to scientifically quantify the impact of an intervention on a particular outcome of interest, are widely employed in a variety of areas. To make informed decisions, technology businesses always conduct A/B testing to evaluate new technologies, using a randomized experiment to compare the performance of each new software implementation with the previous version. Meanwhile, in the medical domain, randomized clinical trials (RCTs) ensuring no systematic differences between treatment groups are the cornerstone of treatment effect evaluation. When analyzing data from REs, analysis of covariance (ANCOVA) is a popular method that can provide consistent results, even if the model is misspecified. REs often require the use of external data to analyze treatment effects better: for example, A/B testing is time-consuming and requires a reasonably high number of users; thus, it is crucial to do a preliminary offline evaluation of external data to implement new interventions more efficiently and eliminate ineffective ones in advance \citep{gilotte2018offline}; meanwhile, if data from earlier clinical stages (Phase I or II) indicate that the product under investigation has a favorable benefit-risk profile in a disease area with unmet healthcare needs, then it is possible to design an RCT with a larger treatment group and a relatively smaller concurrent control (CC) group. Because the small CC group cannot provide sufficient power to the trial, it is reasonable to augment RCTs with external controls (ECs) from earlier trials \citep{yuan2019design}. In this paper, we propose a new method that combines the ECs with CCs to improve average treatment effect (ATE) estimation.

Since \citet{pocock1976combination}, who first introduced historical controls to incorporate external data into analysis, numerous statistical methods have been developed. Specifying a set of covariates in advance and then calculating the propensity score for matching, stratification, or weighting \citep{greenland1999causal,rubin1996matching,rubin2007design,hernan2016using}
is typical. However, these methods assume exchangeability of the conditional outcome distribution given covariates, which is often violated in practice. \cite{FDA2019} identify major sources of bias for ECs, including unmeasured confounding, lack of concurrency, data quality, and outcome validity.
%that there are no unmeasured confounders between ECs and CCs, which is unlikely in real-world applications. 
Additionally, Bayesian approaches \citep{spiegelhalter2004incorporating,hobbs2013adaptive,schmidli2014robust,ibrahim2000power,hobbs2012commensurate,neuenschwander2009note} can handle datasets combining both ECs and CCs: appropriate priors can be selected for incorporating the ECs after evaluating the relationship between the ECs and the CCs. Nevertheless, these methods are dependent on model assumptions, and if the model is misspecified, it can lead to an inflation of type I errors \citep{viele2014use}.
Building upon the work of \citet{stuart2008matching} and \citet{yang2022improved}, who introduced a parametric bias function to adjust for the outcome heterogeneity between the control groups due to unmeasured confounders, \citet{wu2022integrative} advanced the idea by using the sieve estimation approach \citep{chen2007large} to estimate the unknown outcome model and bias function. The bias function in their approach measures the difference between experimental data and observational data. However, \citet{wu2022integrative} did not fully leverage the advantage of REs \citep{wang2021model} since these prior works were based on the assumption that the outcome model was correctly specified. In this paper, we extend the concept of the bias function to handle cases of possible model misspecification. Our bias function measures the difference between the EC outcome mean function and the working model in REs and guarantees consistency even when the outcome mean model is not correctly specified, providing a robust solution to the challenge of model misspecification. This is of great practical significance as simple models such as ANCOVA are commonly used despite the possibility of misspecification. 

We adopt the nonparametric sieve estimation approach \citep{chen2007large} to accurately estimate the bias function, therefore it is important to utilize feature selection techniques. These techniques tackle the high-dimensional aspect of the basis functions used in sieve estimation and address the possibility of the working outcome mean model containing irrelevant covariates. To resolve this, penalized terms can be added to each parameter in the objective function for optimization  along with regularization parameters. However, using the same regularization terms for both the REs and ECs can cause issues due to their differing levels of sparsity and magnitude. Therefore, we implement different penalized terms for the unknown bias function and the working outcome mean model, considering their distinct levels of sparsity and magnitude.

Multiple methods have employed different penalties for different goals. However, they focused on decomposing one function into different parts and applying different penalties to those parts. \citet{chernozhukov2017lava} proposed the Lava estimator by decomposing the signals into a dense part and a sparsity part, and then applying different penalties to each component; \citet{buhlmann2020deconfounding} proposed a spectral deconfounding approach to estimate sparse parameters given hidden variables, and demonstrated the Lava method \citep{chernozhukov2017lava} as one of their special cases; \citet{xing2021adaptive}, further, focused on the estimation of multivariate regression with hidden variables, and demonstrated their method can be viewed as the multivariate generalization of the Lava approach \citep{chernozhukov2017lava}. In addition to decomposing the parameters into sparse and dense parts, \citet{wang2019double} decomposed the function into an easy-to-interpret part and an uninterpretable part, and then applied double penalties. Our approach makes use of double penalties to deal with the possible different structures of the working outcome mean function and bias function to consistently select useful terms and enhance the efficiency of the ATE estimator. 

Different bias functions can utilize varying penalties, depending on their structure. Our study focuses on the use of the Smoothly Clipped Absolute Deviation (SCAD) penalty \citep{fan2001variable,fan2004nonconcave} to illustrate the theorem in the context of variable selection. This is because SCAD offers both oracle properties and asymptotic normality by selecting the appropriate regularization parameters. Nevertheless, the existing results are limited to situations where the models are correctly specified. To overcome this limitation, we present a novel proof for the SCAD penalty that extends its desirable properties to scenarios involving potential misspecification of models.

Our main contributions can be summarized as follows:
\begin{enumerate}
\item We present a novel bias function to combine REs and ECs and use sieve estimation \citep{chen2007large} to provide a flexible and computationally feasible way of estimating the unknown bias function. Our ATE estimator
for REs is consistent regardless of the specification of the working
outcome mean model.
\item We introduce the Double Penalty Integration Estimator (DPIE), which
employs different penalized terms for the unknown bias function and
the working outcome mean model to differentiate their different levels
of sparsity and magnitude. We prove that DPIE guarantees consistency
for the parameters that minimize the least squares loss
%the Kullback-Leibler Information Criterion
% (KLIC) between the true model and the working model \citep{white1982maximum}
and has the oracle property of only selecting non-zero parameters
and exhibiting asymptotic normality under the SCAD penalties.
\item We demonstrate that combining different data sources results in a
more efficient estimated ATE than using only REs.
% as long as the number
% of basis function terms in the bias function is fewer than that of the working outcome mean function. 
The oracle property of DPIE guarantees the selection of relevant basis terms for the bias function, thereby ensuring its precise estimation, which contributes to enhanced efficiency. 
% when the outcome mean function is more complex and less smooth than the bias function,  leading to improved efficiency.
On the other hand, using single penalties may result in a loss of the oracle property and failure to select useful basis terms, leading to decreased efficiency.
%that it can select non-zero basis terms when the outcome mean function
%contains more terms and is less smooth than the bias function, thereby
%improving efficiency. In contrast, single penalties may lose the oracle
%property, fail to select useful basis terms, and result in lost efficiency.
\end{enumerate}
The rest of the paper is organized as follows. We introduce the basic
idea in Section \ref{sec:Problem-Setup}. Section \ref{sec:Method}
introduces the proposed DPIE estimator and derives the theorem. We
conduct simulations for comparison in Section \ref{sec:Simulation}.
Section \ref{sec:Real-Data-Analysis} applies the proposed estimators
to an observational study from the National Supported Work (NSW) and
Current Population Survey (CPS). Finally, we conclude the paper with
a discussion in Section \ref{sec:Discussion}.

\section{PROBLEM SETUP}\label{sec:Problem-Setup}
Denote $X\in\mathcal{X\text{\ensuremath{\subset\mathbb{R}^{d}}}}$ as the vector of pre-treatment covariates, $A\in\text{\{}0,1\text{\}}$ as the binary treatment, and $Y\in\mathbb{R}$ as the outcome of interest. Following the potential outcomes framework \citep{neyman1923application,rubin1974estimating}, let $Y(a)$ be the potential outcome for the subject given the treatment $a,a=0,1.$

In real life, one can use previous trials or real-world data as ECs to supplement the REs. Assuming two data sources are accessible: the RE data source having $n$ independent and identically distributed (i.i.d.) subjects $\left\{ \left(X_{i},A_{i},Y_{i}\right):i\in{\rm \mathcal{I}_{{\rm RE}}}\right\} $ with $n_{1}$ concurrent treatments $\left\{ \left(X_{i},1,Y_{i}\right):i\in{\rm \mathcal{I}_{{\rm CT}}}\right\} $ and $n_{0}$ concurrent controls $\left\{ \left(X_{i},0,Y_{i}\right):i\in{\rm \mathcal{I}_{{\rm CC}}}\right\} $, and the EC data source with $m$ i.i.d. subjects with $\left\{ \left(X_{i},0,Y_{i}\right):i\in{\rm \mathcal{I}_{{\rm EC}}}\right\} $.
Let $N=n+m$ be the total sample size. Define $S$ as the indicator of the subject in the REs: $S_{i}=1$ for $i\in{\rm \mathcal{I}_{{\rm RE}}}$ and $S_{i}=0$ for $i\in{\rm \mathcal{I}_{{\rm EC}}}$. Then the ATE is $\tau=\mathbb{E}\left\{ Y\left(1\right)-Y\left(0\right)\mid S=1\right\} $. Further, let $e\left(X\right)=\mathbb{P}\left(A=1\mid X,S=1\right)$ be the propensity score and also define the conditional outcome mean function as $\mu_{a,s}(X)=\E\left(Y\mid X,A=a,S=s\right )$ %with $\hat{\mu}_{a,s}(X)$ as the corresponding estimator of $\mu_{a,s}(X)$ 
for $a=0,1$ and $s=0,1$.

One of the fundamental challenges to identifying the ATE is that $Y(1)$ and $Y(0)$ cannot be observed simultaneously. To overcome this issue, we make the following three common assumptions in the causal inference literature \citep{rubin1978bayesian}:

\begin{assumption}\label{asump-ignorable} $\{Y(0),Y(1)\}\indep A\mid X,S=1$ almost surely, where $\indep$ means ``independent of''. \end{assumption}
\begin{assumption}\label{consistency} $Y=Y(1)A+Y(0)(1-A)$. \end{assumption}
\begin{assumption}\label{asump-overlap}There exist constants $c_{1}$ and $c_{2}$ such that $0<c_{1}\leq e(X)\leq c_{2}<1$ almost surely.
\end{assumption} 

Assumption \ref{asump-ignorable} states that the treatment assignment is unconfounded in the REs. 
Assumption \ref{consistency} ensures observed outcomes correspond to potential outcomes given the received treatments.
Under Assumptions \ref{asump-ignorable} and \ref{consistency},
%with $X=x$ , we have $\tau(x)=\E\left\{ Y\left(1\right)-Y\left(0\right)\mid X=x,S=1\right\} =\E(Y\mid X=x,A=1,S=1)-\E(Y\mid X=x,A=0,S=1)$, and 
the conditional outcome mean function in REs is $\mu_{a,1}\left(X\right)=\E\left\{ Y\left(a\right)\mid X,S=1\right\} =\E( Y\mid X,A=a,S=1) $.
Assumption \ref{asump-overlap} implies a sufficient overlap of the covariate distribution between the treatment groups, then averaging the treatment effect on the distribution of $X$ is feasible, thus
the ATE is $\tau =\E\left\{ \mu_{1,1}\left(X\right)-\mu_{0,1}\left(X\right)\mid S=1\right\} $.
%$\tau=\E\left\{ \tau(X)\mid S=1\right\} =\E\{ \E(Y\mid X,A=1,S=1)-\E(Y\mid X,A=0,S=1)\mid S=1\} =\E\left\{ \mu_{1,1}\left(X\right)-\mu_{0,1}\left(X\right)\mid S=1\right\} $.

The Analysis of covariance (ANCOVA) model is a powerful tool for estimating the ATE in REs. The randomization design  allows for the ATE estimator $\hat{\tau}$ to be
consistent and asymptotically normal under arbitrary misspecification of the ANCOVA model \citep{wang2021model}. Following the common practice, we use the ANCOVA model as the working model in REs. To enhance the model's generality,
we incorporate a $k_{1}$-dimension basis function of $X$ $p_{\mu}(X)=\left\{ p_{\mu,1}(X),\ldots,p_{\mu,k_{1}}(X)\right\} ^{{\T}}$  into the ANCOVA model as $\bar{\mu}_{A,1}(X;\beta)=\beta_{\rm {int}}+\beta_{A}A+\beta_{X}^{{\T}}p_{\mu}(X)$, where $\beta$ is a $ K_1=(k_1+2)$-dimensional parameter $\beta=\left(\beta_{\rm{int}},\beta_{A},\beta_{X}^{{\T}}\right)^\T$. Under the ANCOVA model, it is common to utilize ordinary least squares estimators for parameter estimation. Denote $\beta_*=\left(\beta_{\rm{int*}},\beta_{A*},\beta_{X*}^{{\T}}\right)^\T$ as 
\begin{equation}\label{betastar}
\beta_*=
\arg\min_{\beta}\E [\{Y-\bar{\mu}_{A,1}(X;\beta)\}^{2} \mid S=1].
\end{equation}
 Importantly, $\beta_{A*}$ is the ATE $\tau$ regardless of the correctness of the working model. 

% Denote $\beta_*=\left(\beta_{\rm{int*}},\beta_{A*},\beta_{X*}^{{\T}}\right)^\T$ as the minimizer 
% of $\E [\{Y-\bar{\mu}_{A,1}(X;\beta)\}^{2} \mid S=1 ]$. 
% %$:=\beta^\T p_\mu$. %, where $\beta^\T=\left(\beta_{\rm{int}},\beta_{A},\beta_{X}^{{\T}}\right)$ and $p_{\mu}^\T=\left\{ 1,A,p_{\mu}^{{\T}}(X)\right\}$ with dimension $K_{1}=k_{1}+2$. 
% Importantly, for $S=1$, $\beta_A$ is the ATE $\tau$ regardless of the correctness of the working model.

To utilize  ECs to augment REs, we consider the following assumption such that the EC covariate distribution is nested in the RE covariate distribution.
\begin{assumption}\label{asump-HCccoverlap}
%$0<\mathbb{P}(S = 1|X) <1$ almost surely.
$f(X\mid S=0)/f(X\mid S=1)<\infty$ almost surely. 
\end{assumption} 

To use ECs to supplement CCs, it is crucial to remove biases of EC data due to possible incomparability between ECs and CCs. We define the bias function as 
\begin{align*}
b_{0}(X)=\mathbb{E}(Y\mid X,A=0,S=0) -\bar{\mu}_{0,1}(X;\beta_*).
\end{align*}
Assumption \ref{asump-HCccoverlap} ensures that $b_0(X)$ is well-defined for all $X$ such that $f(X\mid S=0)>0$. 
If the working model $\bar{\mu}_{0,1}(X;\beta)$ is correctly specified, the bias function reduces to $\mathbb{E}\left( Y\mid X,A=0,S=0\right) -\mathbb{E}\left( Y\mid X,A=0,S=1\right)$, which measures the difference of the conditional mean of the control outcome given $X$ between ECs and CCs. In this case, if the outcome exchangeability  $\E\left( Y\mid X,A=0,S=1\right) =\E\left( Y\mid X,A=0,S=0\right) $ holds, we have $b_{0}(X)\equiv0$; otherwise, $b_{0}(X)\neq0$. This special case is discussed in \citet{wu2022integrative}, but their analysis requires the outcome model to be correctly specified. In contrast, our setup does not necessitate a correctly specified outcome model. 

For the combined data, let the ANCOVA model augmented by the  bias function $b_{0}(X)$ be $$\bar{\mu}_{A,S}(X;\beta)=\beta_{\rm{int}}+\beta_{A}A+\beta_{X}^{{\T}}p_{\mu}(X)+(1-S)b_{0}(X),$$ then
\begin{align*}
&\bar{\mu}_{0,0}(X;\beta)=\beta_{\rm{int}}+\beta_{X}^{{\T}}p_{\mu}(X)+b_{0}(X)\\
  =&\bar{\mu}_{0,1}(X;\beta)+\mathbb{E}(Y\mid X,A=0,S=0)-\bar{\mu}_{0,1}(X;\beta)\\
  =&\mathbb{E}(Y\mid X,A=0,S=0).
\end{align*}
An important implication is that even if the outcome working model is misspecified, incorporating the bias function $b_{0}(X)$ ensures that $\bar{\mu}_{0,0}(X;\beta)$ recovers the true outcome mean under ECs. 
%This helps to account for any unmeasured confounding factors and ensures that the estimated treatment effect $\hat{\tau}$ is consistent and can be more efficient by leveraging additional information in ECs. 
 
The following theorem demonstrates that  $\beta_{A*}$ still identifies the ATE $\tau$ in the combined RE and EC data.

\begin{theorem} \label{iden}(Identification)  
%The ANCOVA model is
%\[
%Y=\beta_{\rm{int}}+\beta_{A}A+\beta_{X}^{{\T}}p_{\mu}(X)+(1-S)b_{0}(X)+\epsilon,\ \E(\epsilon)=0.
%\]
Under Assumptions \ref{iden}--\ref{asump-HCccoverlap} and the augmented ANCOVA model, $\beta_*=\left(\beta_{\rm{int*}},\beta_{A*},\beta_{X*}^{{\T}}\right)^\T$  defined in \eqref{betastar} minimizes  $\E [\{Y-\bar{\mu}_{A,S}(X;\beta)\}^{2}]$, and
$\beta_{A*}=\tau.$ 
\end{theorem}

Theorem \ref{iden} provides a vehicle to integrate REs and ECs for robust estimation of the ATE.  Consistent estimation of ${\tau}$ still depends on an accurate approximation  of unknown $b_{0}(X).$ 
Thus, we adopt the method of sieves \citep{chen2007large}. Denote $p_{b}(X)$ as the $K_{2}$-dimensional basis functions. Based on Theorem S1 in the Supplementary Material, there exists a $K_{2}$-vector $\delta_{*}$
such that the uniform convergence 
$\text{\ensuremath{p_{b}^{{\T}}(X)\delta_{*}\rightarrow}} b_{0}(X)$ and therefore the uniform convergence 
\begin{align*}
&\beta_{\rm{int}}+\beta_{A}A+\beta_{X}^{{\T}}p_{\mu}(X)+(1-S)p_{b}^{{\T}}(X)\delta_{*}\\
\rightarrow & \beta_{\rm{int}}+\beta_{A}A+\beta_{X}^{{\T}}p_{\mu}(X)+(1-S)b_{0}(X)
\end{align*} 
hold as $K_{2}\rightarrow\infty$. 
Then our final working model becomes 
\begin{equation*}
\bar\mu_{A,S}(X;\beta,\delta) =\beta_{\rm{int}}+\beta_{A}A+\beta_{X}^{{\T}}p_{\mu}(X)+(1-S)\delta^{{\T}}p_{b}(X).
\end{equation*}
We consider using the least squares loss function to obtain estimators for $\beta_*$ and $\delta_*$. 
To overcome the risk of overfitting in sieve estimation, where high-dimensional
basis functions are used, it is necessary to add regularizers on $\delta_*$ %$\Lambda_{\delta}$
to the loss function. Additionally, as the working models $\bar{\mu}_{A,1}(X;\beta)$
may contain irrelevant covariates, adding regularizers on $\beta_*$ %$\Lambda_{\beta}$ 
is recommended to select proper covariates. 
In the subsequent section, we thoroughly explore the significance of the structural properties of regularizers to effectively accommodate the inherent characteristics of the problem at hand.

\section{A DOUBLE PENALTY REGULARIZATION METHOD}\label{sec:Method}
\subsection{Main Idea}\label{subsec:Main-Idea}

The common regularization methods use the same regularization parameter %$\lambda$ 
for the two penalty functions on $\beta_*$ and $\delta_*$. %$\Lambda_{\beta,\lambda}$ and $\Lambda_{\delta,\lambda}$. 
However, it is important to note that $\beta_{*}$ and $\delta_{*}$ may have distinct complexities (magnitude and/or sparsity), and it is beneficial to apply different penalties, instead of the same penalty, to both parameters.

To begin with, we denote $P_{\lambda}\left(\gamma\right)=\lambda P(\gamma)$
as the penalty function with regularization parameter $\lambda$ for
any parameter $\gamma$ and any penalization $P(\cdot)$. 
There are various choices for the penalty function, like Lasso \citep{tibshirani1996regression},
Smoothly Clipped Absolute Deviation (SCAD) \citep{fan2001variable,fan2004nonconcave}
penalties, or use black-box methods like the random forest. 
To overcome the limitations caused by adding the same penalty to all
parameters, we set up different penalties for $\beta$ and $\delta$ and propose a double penalty-regularized integration estimator (DPIE). Specifically, the proposed DPIE under the least squares loss is 
\begin{equation*}
\begin{split}
\left(\hat{\beta},\hat{\delta}\right)&=\underset{\beta,\delta}{\rm argmin} 
\Biggl[
\sum_{i=1}^{N}\{Y_i-\bar{\mu}_{A_i,S_i}(X_i;\beta,\delta)\}^{2}\\
&+N\sum_{j=1}^{K_{1}}P_{\lambda_{1,j}}\left(\vert\beta\vert\right)+N\sum_{j=1}^{K_{2}}P_{\lambda_{2,j}}\left(\vert\delta\vert\right)
\Biggl].
\end{split}
\end{equation*}

To implement regularization, selecting the two regularization parameters, $\lambda_{1}$ and $\lambda_{2}$, is important. There are multiple methods to search for these parameters. One simple way is to define a scaling parameter $sc$ as $sc=\lambda_{2}/\lambda_{1},$ then one can use cross-validation to choose $sc$ given a particular search range and within each $sc$
value, one can also use cross-validation to choose $\lambda_{2}.$ Both cross-validation steps can use the software $\texttt{cv.ncvreg}$ function in the R package $\texttt{ncvreg},$ which finds the regularization parameter based on the minimum cross-validated error.

If we only use the RE data, we have 
\begin{equation*}
\begin{split}
\hat{\beta}_{\rm {RE}}&=\underset{\beta}{\rm argmin} 
\Biggl[
\sum_{i=1}^{N} S_{i} \{Y_i-\bar{\mu}_{A_i,S_i}(X_i;\beta,\delta)\}^{2}\\&+n\sum_{j=1}^{K_{1}}P_{\lambda_{1,j}}\left(\vert\beta\vert\right)
\Biggl].
\end{split}
\end{equation*}

\subsection{Theoretical Properties}\label{subsec:Theoretical-Analysis}

The goal of this subsection is to derive the statistical properties of the DPIE. More interestingly, we aim to show the DPIE gains efficiency over the RE-only estimator of the ATE. 

For concreteness, we will focus on using the SCAD penalty for illustration. It is worth mentioning that our setup can also be developed for different penalty functions, but we will not
be discussing those in this study.  
When the working function $\bar\mu_{A,S}(X;\beta,\delta)$ is correctly specified, % and the  density function of $(Y,A,S,X)$ is known, 
the penalized maximum likelihood estimator with the SCAD penalty has both oracle properties as well as asymptotic normality by
selecting the appropriate regularization parameter $\lambda$ \citep{fan2001variable,fan2004nonconcave}.  
In the following, we will show the DPIE derived using the penalized least squares loss function with double SCAD penalties maintains the oracle property and asymptotic normality under the working model $\bar\mu_{A,S}(X;\beta,\delta)$. 

Following the framework in \citet{fan2001variable,fan2004nonconcave}, we rewrite the working model as
\begin{align*}
&\bar\mu_{A,S}(X;\beta,\delta)\\=&\beta_{\rm{int}}+\beta_{A}A+\beta_{X}^{{\T}}p_{\mu}(X)+(1-S)\delta^{{\T}}p_{b}(X)\\=&p^{\T}\theta,
\end{align*}
where $\theta^{{\T}}=(\beta^{{\T}},\delta^{{\T}})$ and $p=\{ 1,A,p_{\mu}^{{\T}}(X),(1-S)p_{b}^{{\T}}(X)\} ^{{\T}}$ with dimension $K=K_{1}+K_{2}$. 
%where $K_{1}$ and $K_{2}$ increase with sample size $N$ such that 
Further, denote $p_\mu=\{1,A,p_\mu(X)^{\rm{T}}\}^{\rm{T}}$ and $p_b=(1-S)p_b(X)$, therefore
in REs, we have $p^{{\T}}\theta=p_{\mu}^{{\T}}\beta$, and in ECs, we have $p^{{\T}}\theta=p_{\mu}^{{\T}}\beta+p_{b}^{{\T}}\delta.$ 
Denote $g$ as the unknown true density function of $(Y,p)$ and $f$ as the working density function such that minimizing the least squares loss is equivalent to maximizing the quasi-log-likelihood function; the Gaussian distribution is one such example. Denote $P_{\lambda}\left(\theta\right)$
as the SCAD penalty function with the first-order derivative 
\[
P_{\lambda}^{'}(\theta)=\lambda\left\{ \bone\left(\theta\leq\lambda\right)+\frac{(a\lambda-\theta)_{+}}{\left(a-1\right)\lambda}\bone\left(\theta>\lambda\right)\right\} 
\]
for some $a>2$ and $\theta>0$. 
%The goal of the theoretical analysis part is to guarantee the consistency
%and the oracle property of the parameters $\beta_{*},\delta_{*}$,
%further, we also show using the combined data is at least as efficient
%as using only the REs. 
Then one can rewrite the penalized least squares estimator as 
the penalized quasi-likelihood estimator 
\begin{equation*}
\begin{split}
\left(\hat{\beta},\hat{\delta}\right)&=\underset{\beta,\delta}{\rm argmax} \;Q(\beta,\delta)\\
&=\underset{\beta,\delta}{\rm argmax} 
\Biggl[
\sum_{i=1}^{N}\left\{ \ln f\left(Y_{i},p_{i,}\beta,\delta\right)\right\}\\
&-N\sum_{j=1}^{K_{1}}P_{\lambda_{1,j}}\left(\vert\beta\vert\right)-N\sum_{j=1}^{K_{2}}P_{\lambda_{2,j}}\left(\vert\delta\vert\right)
\Biggl].
\end{split}
\end{equation*}

Denote $\beta_{*}=(\beta_{*,1},\ldots,\beta_{*,K_1})^{\rm{T}}$ and $\delta_{*}=(\delta_{*,1},\ldots,\delta_{*,K_2})^{\rm{T}}$.
Let 
\begin{align*}
\alpha_{N} & =\max_{1\leq j_{1}\leq K_{1},1\leq j_{2}\leq K_{2}}\left\{ 
\begin{aligned}
&P_{\lambda_{1}}^{'}\left(\vert\beta_{*,j_{1}}\vert\right),P_{\lambda_{2}}^{'}\left(\vert\delta_{*,j_{2}}\vert\right),\\&\beta_{*,j_{1}}\neq0,\delta_{*,j_{2}}\neq0
\end{aligned}
\right\} ,\\
b_{N} & =\max_{1\leq j_{1}\leq K_{1},1\leq j_{2}\leq K_{2}}\left\{ 
\begin{aligned}
& P_{\lambda_{1}}^{''}\left(\vert\beta_{*,j_{1}}\vert\right),
P_{\lambda_{2}}^{''}\left(\vert\delta_{*,j_{2}}\vert\right),\\&\beta_{*,j_{1}}\neq0,\delta_{*,j_{2}}\neq0
\end{aligned}
\right\} ,
\end{align*}
where $P_{\lambda}^{''}\left(\theta\right)$ is the second-order derivative
of $P_{\lambda}(\theta).$ 

We impose the regularity conditions similar to \citet{fan2004nonconcave}.

\begin{assumption}\label{assumption3}
%Let the values of $\beta_{*,1},\ldots,\beta_{*,s_{1}}$
% be nonzero and $\beta_{*,s_{1}+1},\ldots,\beta_{*,K_{1}}$ be zero.
% Similarly, let the values of $\delta_{*,1},\ldots,\delta_{*,s_{2}}$
% be nonzero and $\delta_{*,s_{2}+1},\ldots,\delta_{*,K_{2}}$ be zero.
The parameter values $(\beta_{*},\delta_{*})$ and regularization parameters $(\lambda_1,\lambda_2)$ satisfy 
\begin{align*}
\min_{1\leq j\leq s_{1}}\vert\beta_{*,j}\vert/\lambda_{1}\rightarrow\infty,\  & \min_{1\leq j\leq s_{2}}\vert\delta_{*,j}\vert/\lambda_{2}\rightarrow\infty,\\
\max_{s_{1}+1\leq j\leq K_{1}}\vert\beta_{*,j}\vert/\lambda_{1}\rightarrow0,\  & \max_{s_{2}+1\leq j\leq K_{2}}\vert\delta_{*,j}\vert/\lambda_{2}\rightarrow0,
\end{align*}

as $N\rightarrow\infty.$

\end{assumption}

Assumption \ref{assumption3} divides $\beta_*$ and $\delta_*$ into non-zero segments, i.e., $s_1$-dimensional $\beta_{*1}$ and $s_2$-dimensional $\delta_{*1}$, and negligible segments, i.e., $(K_1-s_1)$-dimensional $\beta_{*2}$ and $(K_2-s_2)$-dimensional $\delta_{*2}$. If $\beta_{*2}$ and $\delta_{*2}$ are exactly zero, they satisfy the above conditions automatically. 
We then denote $\theta_{*}$ as $(\theta_{*1}^{\T}, \theta_{*2}^{\T})^{\T}$, with $\theta_{*1}=(\beta_{*1}^{\rm{T}},\delta_{*1}^{\rm{T}})^{\rm{T}}$ of dimension $s=s_1+s_2$, and $\theta_{*2}=(\beta_{*2}^{\rm{T}},\delta_{*2}^{\rm{T}})^{\rm{T}}$.

% $\beta_*=(\beta_{*1}^{\T},\beta_{*2}^{\T})^{\T}$ and $\delta_{*}=(\delta_{*1}^{\T}, \delta_{*2}^{\T})^{\T}$, $\theta_{*}=(\theta_{*1}^{\T}$

% Denote $\beta_*=(\beta_{*1}^{\T},\beta_{*2}^{\T})^{\T}$, where $\beta_{*1}\neq0$ with dimension $s_{1}$ and $\beta_{*2}=0$.
% Similarly, denote  $\delta_{*}=(\delta_{*1}^{\T}, \delta_{*2}^{\T})^{\T}$, $\theta_{*}=(\theta_{*1}^{\T}, \theta_{*2}^{\T})^{\T}$
% where $\delta_{*1}\neq0$ with dimension $s_{2}$ and $\delta_{*2}=0,$
% and $\theta_{*1}\neq0$ with dimension $s=s_{1}+s_{2}$  and $\theta_{*2}=0$.

\citet{fan2004nonconcave} showed under Assumption \ref{assumption3}, for the SCAD penalties, $\alpha_{N}=0$ and $b_{N}=0$ as $N$ is large enough, where the former 
% guarantees the unbiasedness property in the
% asymptotic normality property and 
ensures the existence of root-$N/K$-consistent
penalized likelihood estimator, and the latter ensures the penalty function does not have much more influence on the penalized likelihood
functions, making the penalty estimator have the same efficiency as
the maximum likelihood estimator.  
% Assumption S10 ensures that, upon selecting an appropriate working density function $f$, the OLS estimator will converge to the value which minimizes the $KLIC$ distance between the working density function $f$ and the true density function $g$.
%Regularity conditions S1--S8 on
%likelihood function and density functions are in the Supplementary Material. 

Denote $\|v\|_{p}$
as the $\mathcal{L}_{p}$-norm of a vector $v$. Based on these assumptions,
we can provide the consistency and the asymptotic normality of the
estimated parameters.
\begin{theorem}\label{theorem1}Suppose that the working density function $f\left(Y,p,\beta,\delta\right)$ and the true density function $g(p,Y)$
satisfy Assumptions S1--S9 on the Supplementary Material, and the
SCAD penalty functions $P_{\lambda_{1}}\left(\cdot\right),P_{\lambda_{2}}\left(\cdot\right)$
satisfy Assumption \ref{assumption3}. If $K^{4}/N\rightarrow0$ as
$N\rightarrow\infty$, then there is a local maximizer $\left(\hat{\beta},\hat{\delta}\right)$
of $Q(\beta,\delta)$ such that $\|\hat{\beta}-\beta_{*}\|_{2}=O_{p}\left\{ \left(K/N\right)^{1/2}\right\} ,\|\hat{\delta}-\delta_{*}\|_{2}=O_{p}\left\{ \left(K/N\right)^{1/2}\right\} .$

\end{theorem}

% Denote 
% \begin{align*}
% &A\left(\theta\right)=-\E\left[\frac{\partial^{2}\log f\left(Y_{1},p_{1},\theta\right)}{\partial\theta_{j}\partial\theta_{k}}\right],\\
% &B(\theta)=\E\left[\left\{ \frac{\partial\log f\left(Y_{1},p_{1},\theta\right)}{\partial\theta}\right\} \left\{ \frac{\partial\log f\left(Y_{1},p_{1},\theta\right)}{\partial\theta}\right\} ^{{\T}}\right],
% \end{align*}
% and also denote $I(\theta_{0})$ be the Fisher information matrix,
% and let $I(\theta_{01})=I(\theta_{01},0)$ be the Fisher information
% matrix knowing $\theta_{02}=0$. Note, if the density function $f$
% is correctly specified, then $A(\theta_{0})=B(\theta_{0})=I(\theta_{0}).$
Then we have the following theorem. 
\begin{theorem}\label{theorem2}
Under Assumption \ref{assumption3} and Assumptions S1--S9 in the
Supplementary Material, if $\lambda_{1},\lambda_{2}\rightarrow0$,
$\sqrt{N/K}\lambda_{1}\rightarrow\infty,\sqrt{N/K}\lambda_{2}\rightarrow\infty$
and $K^{5}/N\rightarrow0$ as $N\rightarrow\infty$, then with probability
tending to 1, $\hat{\beta},\hat{\delta}$ in Theorem \ref{theorem1}
must satisfy 
\begin{enumerate}
\item (Sparsity) $\hat{\beta}_{2}=0,\hat{\delta}_{2}=0$. 
\item (Asymptotic normality) 
\begin{align*}
&\sqrt{N}WA^{1/2}\left(\theta_{*1}\right)\left(\hat{\theta}_{1}-\theta_{*1}\right)\\
&\rightarrow\mathcal{N}\left(0,WA^{-1/2}(\theta_{*1})B(\theta_{*1})A^{-1/2}(\theta_{*1})W^{{\T}}\right)
\end{align*}
in distribution, where $W$ is a $q\times s$ matrix such that $WW^{{\T}}\rightarrow G$,
and $G$ is a $q\times q$ nonnegative symmetric matrix. For simplicity, the specific forms of $W$, $A(\theta)$ and 
 $B(\theta)$ are deferred to the Supplementary Material.
\end{enumerate}
\end{theorem}
Theorems \ref{theorem1}-\ref{theorem2} demonstrate
that under proper selection of regularization parameters, the estimator $\hat{\theta}$ is consistent for $\theta_{*}$ and asymptotic normal. % If the model
% is specified correctly, i.e., $g(Y,p_{i})=f(Y,p_{i},\theta)$ for
% some $\theta\in\Theta$, then $\theta_{0}=\theta_{*}$, and 
% \[\sqrt{N}WI^{1/2}\left(\theta_{01}\right)\left(\hat{\theta}_{1}-\theta_{01}\right)\rightarrow\mathcal{N}\left(0,WW^{{\T}}\right)
% \]
% in distribution. 

On the other hand, using a single $\lambda$ may not yield the desired results as stated in Theorem \ref{theorem2}. This can be illustrated by a simple analytical calculation demonstrating that a single $\lambda$ does not satisfy Assumption \ref{assumption3}.
Let $\beta_{*1}=O_{p}(N^{-1/2}),\beta_{*2}=O_{p}(N^{-1}),\delta_{*1}=O_{p}\left(N^{-1/10}\right),\delta_{*2}=O_{p}\left(N^{-1/3}\right).$
Assuming $\lambda_{1}=N^{\epsilon}$ and $\lambda_{2}=N^{\gamma}$
such that $\beta_{*1},\beta_{*2},\delta_{*1}$ and $\delta_{*2}$ satisfy Assumption \ref{assumption3}: 
\begin{align*}
\frac{N^{-1/2}}{N^{\epsilon}}\rightarrow\infty,\  & 
\frac{N^{-1/10}}{N^{\gamma}}\rightarrow\infty,\ 
\\
\frac{N^{-1}}{N^{\epsilon}}\rightarrow0,\  & \frac{N^{-1/3}}{N^{\gamma}}\rightarrow0.
\end{align*}
Then, we should have $-1<\epsilon<-1/2$ and $-1/3<\gamma<-1/10$, which
means it is impossible to reduce $\lambda_1=N^\epsilon$ and $\lambda_2=N^\gamma$ to a single $\lambda$ that satisfies the conditions in Assumption \ref{assumption3}.
This simple example shows that if the magnitude of $\beta_*$ and $\delta_*$ differ largely, one single penalty cannot satisfy the requirements for consistency and oracle properties. 
A toy numerical experiment in the Supplemental Materials demonstrates that utilizing different penalties for $\beta_{*}$ and $\delta_{*}$ yields superior performance compared to using a single penalty.
The simulation study in Section \ref{sec:Simulation} corroborates this phenomenon. 

It should be noted that Theorem \ref{theorem1}-\ref{theorem2} can be expanded to incorporate different working density functions $f$, not just the current form corresponding to the least squares loss. This means we can also adopt other losses beyond the least squares loss. If $f$ takes other forms, under certain regularity conditions on $f$ and $g$, $\hat\theta$ that maximizes the penalized quasi-log-likelihood function $Q(\theta)$ converges to $\tilde\theta_*$, where $\tilde\theta_*$ minimizes the Kullback-Leibler Information Criterion (KLIC) between $f$ and $g$, $KLIC(g:f,\theta)=\mathbb{E}\left[\log\left\{ g(Y,p)/f(Y,p,\theta)\right\} \right]$. %Detailed discussions are included in the Supplementary Material. 
However, the identification strategy for $\tau$ from $f(Y,p,\tilde\theta_*)$ may change, depending on the specific form of $f(Y,p,\tilde\theta_*)$, similar to \cite{wang2021model}. Future work will focus on extending the DPIE of $\tau$ to other loss functions.

% Theorem \ref{theorem2}  establishes asymptotic normality for a general estimate $\hat\beta$ that converges to $\beta_*$, where $\beta_*$ in conjunction with $\delta_*$ minimizes the $KLIC$ distance between working density $f$ and the true density $g$. In the ensuing Theorem, our focus is on the OLS estimate $\hat\beta$, demonstrating that under the OLS corresponding working density $f$, $\beta_{A*}$ equals $\tau$.

\subsection{Comparison between the DPIE and the RE-only estimator}
We now show the advantage of the DPIE of the ATE $\hat\tau=\hat\beta_A$ based on the combined data compared with the RE-only estimator $\hat\tau_{{\rm RE}}=\hat\beta_{A,{\rm RE}}$. For ease of comparison, we assume homoscedasticity of the residual error $\epsilon$ in the equation $Y=\bar\mu_{A,S}(X)+\epsilon$, where $\V(\epsilon)=\sigma^{2}$.

\begin{theorem}\label{theorem 4}
Under Assumption \ref{iden}--\ref{assumption3} and Assumptions S1--S9 in
Supplementary Material, if $\lambda_{1},\lambda_{2}\rightarrow0$,
$\sqrt{N/K}\lambda_{1}\rightarrow\infty,\sqrt{N/K}\lambda_{2}\rightarrow\infty$
and $K^{5}/N\rightarrow0$ as $N\rightarrow\infty$. For the penalized least squares estimators, 
%     \item
% $\delta_{*}=\delta_{0},\beta_{*}=\left(\beta_{0*},\beta_{A*},\beta_{X*}^{{\T}}\right)^{{\T}}$ satisfies $\beta_{A*}=\tau.$
\begin{enumerate}
    \item in the combined RE and EC data, 
$\sqrt{N}(\hat{\tau}-\tau)\rightarrow\mathcal{N}\left\{ 0,\V(\hat{\tau})\right\} $ in distribution;
\item in the RE data, 
$\sqrt{n}(\hat{\tau}_{{\rm RE}}-\tau)\rightarrow\mathcal{N}\left\{ 0,\V(\hat{\tau}_{{\rm RE}})\right\} $
 in distribution; and
 \item $\V(\hat{\tau}_{{\rm RE}})>\V(\hat{\tau})$.
%  \item $\V(\hat{\tau}_{{\rm RE}})\geq\V(\hat{\tau})$
% and the equality holds iff $p_{\mu}=Mp_{b}(X)$ for some matrix
% $M$.
\end{enumerate}
% the combined data has
% $\hat{\tau}-\tau\rightarrow\mathcal{N}\left\{ 0,\V(\hat{\tau})\right\} $ in distribution, and the RE data has 
% $\hat{\tau}_{{\rm RE}}-\tau\rightarrow\mathcal{N}\left\{ 0,\V(\hat{\tau}_{{\rm RE}})\right\} $
%  in distribution, where $\V(\hat{\tau}_{{\rm RE}})\geq\V(\hat{\tau})$
% and the inequality holds iff $p_{\mu}=Mp_{b}(X)$ for some matrix
% $M$.

\end{theorem}
Theorem \ref{theorem 4} shows that the estimate of $\tau$ will remain
accurate, regardless of the validity of the ANCOVA working model.
When incorporating ECs, the DPIE estimator gains efficiency over the RE-only estimator. 
%Additionally, adding double penalties in SCAD ensures the accurate estimation of $\delta_{*}$, which in turn guarantees the accuracy of the bias function. As demonstrated in Theorem 1, having an accurate bias function is essential for achieving consistency in the ATE estimation, even when the ANCOVA working model is misspecified. This highlights the significance of using different penalties for the parameters $\beta_{*}$ and $\delta_{*}$. 

\section{SIMULATION}\label{sec:Simulation}

In this section, we demonstrate the importance of applying distinct penalties for different data sources. Subsequently, we compare the proposed estimator of $\tau$ with the RE-only estimator and existing competitors that combine REs and ECs.

\subsection{Simulation Study 1}

We generate two data sources with sample sizes $n=m=1000$. % and total sample size  $N=2000$. 
Covariates $X\in\mathbb{R}^{50}$ are generated by $X_{d}\sim{\rm Uniform}\left[1-\sqrt{3},1+\sqrt{3}\right],d=1,\ldots,50$, and 
outcome is generated by $Y=X^{{\T}}\beta_{0}+\left(1-S\right)X^{{\T}}\delta_{0}+\epsilon,$
where $\epsilon\sim\mathcal{N}\left(0,1\right)$. We simulate $T=100$
Monte Carlo times, and specify the true $\beta_{0}=(1,\ldots,50)^{{\T}}/50.$

To examine the instances where various penalties are required and validate the argument presented in Section \ref{sec:Method}, we specify three cases for $\delta_{0}$: 
\begin{enumerate}
\item $\|\delta_{0}\|_{1}\geq \|\beta_{0}\|_{1}$ and half of parameters in $\delta_{0}$ equal to zero: $c\|\beta_{0}\|_{1}=\|\delta_{0}\|_{1}$ and $c=1,3,5,7,9.$ 
\item $\|\delta_{0}\|_{1}<\|\beta_{0}\|_{1}$ and half of parameters
in $\delta_{0}$ equal to zero: $c\|\beta_{0}\|_{1}=\|\delta_{0}\|_{1}$ and $c=0.1,0.3,0.5,0.7,0.9.$ 
\item Vary the sparsity level of $\delta_{0}$ while ensuring that its magnitude satisfies $\|\delta_{0}\|_{1}=\|\beta_{0}\|_{1}$:
%the number of variables in $\delta_0$ equal to zero one by one. 
the number of variables in $\delta_0$ that equal zero gradually change, increasing from $2$ to $50$ with a step size of three.
\end{enumerate}
In each case, we compare the following estimators:
\begin{enumerate}
\item  the proposed estimator based on the combined data using the double
SCAD penalty (denoted as
``DPIE''), 
\item the estimator based on the combined data using the single
SCAD penalty (denoted as ``SPIE''), and
\item the estimator based only on the RE data (denoted as ``RE''). 
\end{enumerate}
\begin{figure*}
\center{}\includegraphics[scale=0.25]{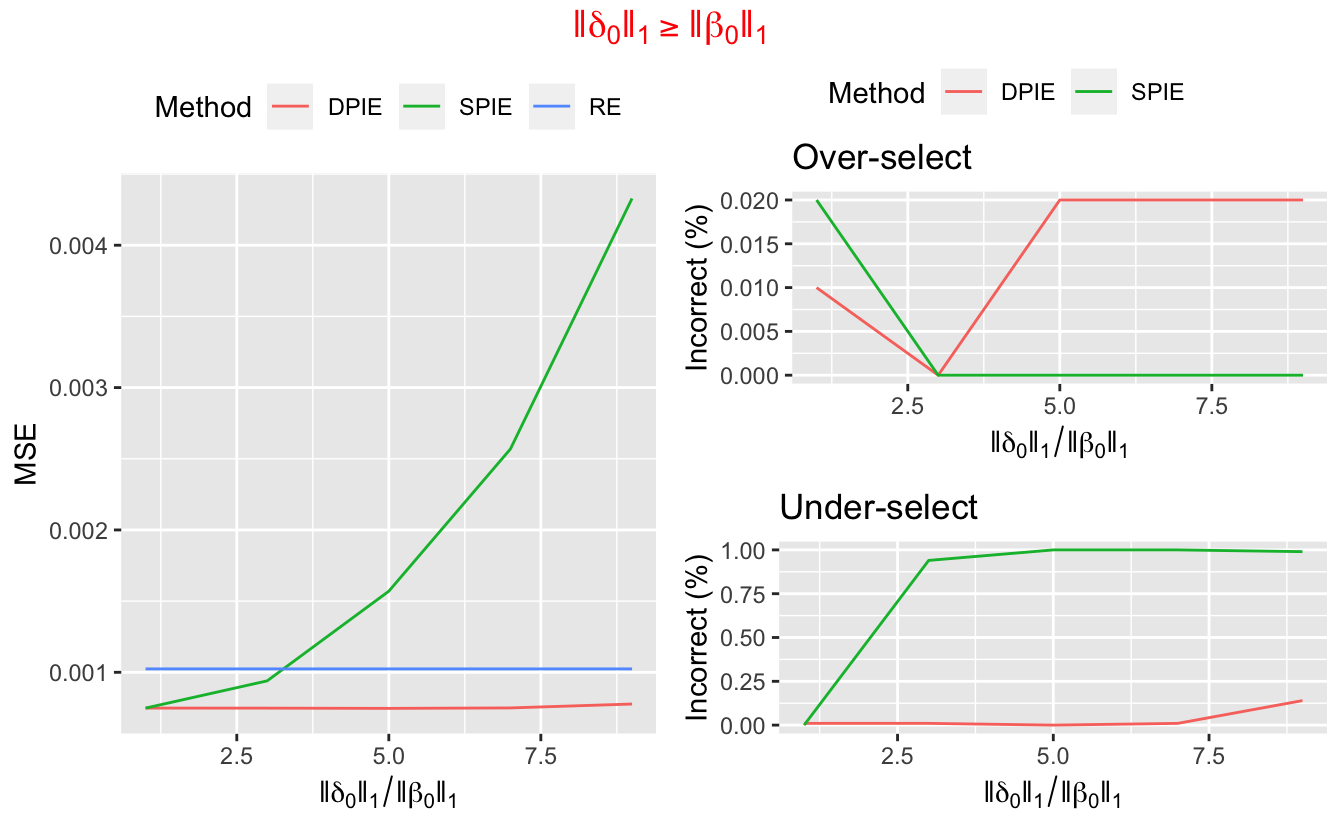}
\caption{\label{large}Simulation results based on 100 Monte Carlo times. The
left panel shows the MSE versus the magnitude ratio between $\delta_{0}$
and $\beta_{0}$. The right panel shows the percentage of wrongly
choosing more and fewer parameters, separately.}
\end{figure*}
All results are based on re-fitting models with the parameters chosen
in each method, and compare the results based on the mean squared
error $MSE=\sqrt{d^{-1}\sum_{i=1}^{d}\left(\hat{\beta}_{i}-\beta_{0,i}\right)^{2}}$
and the percentage of incorrectly selecting more (denoted as ``Over-select'') and fewer parameters (denoted as ``Under-select''). Figure \ref{large} shows the MSE results and the percentage of Under-select and Over-select in case a).
When the magnitudes of two parameters differ, using different penalties
improves accuracy when compared to using the same penalties for all
parameters. Moreover, the gained accuracy improves as the magnitude
difference increases. The right panel of Figure \ref{large} shows
the percentage of incorrectly selecting more or fewer variables. When $\|\delta_{0}\|_{1}\geq\|\beta_{0}\|_{1}$,
using the same penalties makes it difficult to select $\beta_{0}$, resulting in a large MSE. These findings are consistent
with the theoretical results in Section \ref{sec:Method}. 
Case b) for $\|\delta_{0}\|_{1}<\|\beta_{0}\|_{1}$ shows a similar phenomenon, and thus the results are deferred to the Supplementary
Material.

In contrast, in Case c), where we vary the sparsity levels of $\delta_{0}$ while keeping the magnitude the same ($||\delta_{0}||_1=||\beta_{0}||_1$), the DPIE and SPIE methods demonstrate similar performances. Refer to the figure in the Supplementary Material for a visual representation. This finding also aligns with the theoretical result in
Section \ref{sec:Method}, where we only need to restrict the magnitude
of different parameters to guarantee consistency and oracle properties.

\subsection{Simulation Study 2}

We now compare the DPIE of $\tau$ with the RE-only estimator and existing competitors combining the REs and the ECs. 
We generate REs and ECs with sample sizes $n=m=1000$. 
%We set up $T=500$ simulation times. Let the total sample size $N=2000,$
%where the RE sample size $n=1000$, and the EC sample size $m=1000$.
Covariates $X\in\mathbb{R}^{2}$ are generated by $X_{d}\sim{\rm Uniform}\left[-1.5,1.5\right],d=1,2$.
The treatments $A$ in the REs are generated by ${\rm Binomial}(1000,0.5)$.
We consider two settings for generating outcomes: 
\begin{enumerate}
\item[S1.]  $Y=-1.5X_{1}^{2}-1.5X_{2}+2A+(1-S)(10X_{1}^{2}+4X_{2}^{3})+\epsilon$,
where $\text{\ensuremath{\epsilon\sim\mathcal{N}(0,1)}}$; 
\item[S2.]  $Y=-1.5X_{1}^{2}-1.5e^{X_{2}}+2A+(1-S)(10X_{1}^{2}+4X_{2}^{3})+\epsilon$,
where $\text{\ensuremath{\epsilon\sim\mathcal{N}(0,1)}}$. 
\end{enumerate}
In each case, we specify $\bar{\mu}_{1,1}(X;\beta),\ \bar{\mu}_{0,1}(X;\beta)$
and $b_{0}(X)$ by using the power series basis functions with a maximum power of three. We use the double  SCAD penalty method to select
important features of $\bar{\mu}_{1,1}(X;\beta),\ \bar{\mu}_{0,1}(X;\beta)$ and
$b_{0}(X)$, where, in Setting S1, the working models are correct,
while in Setting S2, the working models are misspecified for $\bar{\mu}_{1,1}(X;\beta),\ \bar{\mu}_{0,1}(X;\beta)$.
After selecting parameters, we estimate the variance of $\hat{\tau}$
using standard software output from linear regression. We
compare our method with the Bayesian Power Prior (BPP) method \citep{lin2019propensity} and the matching and bias adjustment (MBA) procedure \citep{stuart2008matching}.

The BPP method leverages the estimated probability of trial inclusion $\mathbb{P}(S=1\mid X)$ as the power to downweight the contribution of the likelihood of each EC subject. It requires the full
likelihood function to be correctly specified. To meet this requirement, we utilize power series basis functions with a maximum power of three to fit the likelihood functions for both REs and ECs. This grants more flexibility to fulfill the rigorous assumption.
The MBA approach consists of three stages: in stage 1, match concurrent treated units (CTs) with CCs, leaving some treatment units potentially unmatched; in stage 2, match CCs with ECs to measure the bias $\delta$ between ECs and CCs; in stage 3, match the unmatched CTs in the first stage with ECs, adjusting the matched ECs by removing the bias term $\delta$. The MBA average treatment effect is the standard treatment effect analysis applied to stage 1 matched CT and CC pairs and stage 3 matched CT and EC pairs adjusting for the bias term.

\begin{table}[ht]
\centering{} \caption{\label{table1}The absolute bias, true variance, MSE, estimated variance, and the 95\% Wald confidence intervals of  DIPE, ANCOVA, BPP (Bayesian Power Prior), MBA (Matching and Bias Adjustment) in two settings (denoted as S1, S2). All numbers are the true values times $10^{-3}$ except for CI. The results with the best performance are highlighted in bold. }
\begin{tabular}{ccccc}
\hline 
 & RE & \multicolumn{3}{c}{EC+RE}\tabularnewline
\multicolumn{2}{r}{ANCOVA} & DIPE & BPP & MBA\tabularnewline
\hline 
\multicolumn{5}{c}{S1}\tabularnewline
\hline 
$\vert\hat{\tau}-\tau\vert$ & 4.05 & \textbf{2.43} & 110.7 & 646 \tabularnewline
$v$ & 3.94 & \textbf{3.33} & 326 & 457 \tabularnewline
MSE & 3.96 & \textbf{3.34} & 338 & 874\tabularnewline
$\hat{v}$ & 4.00 & 3.27 & -- & --\tabularnewline
CI & 95.6\% & 94.6\% & -- & --\tabularnewline
\hline 
\multicolumn{5}{c}{S2}\tabularnewline
\hline 
$\vert\hat{\tau}-\tau\vert$ & 4.67 & \textbf{4.42} & 111.5  & 645\tabularnewline
$v$ & 3.97 & \textbf{3.33} & 328  & 457 \tabularnewline
MSE & 3.99 & \textbf{3.35} & 340 & 873\tabularnewline
$\hat{v}$ & 4.01 & 3.30 & -- & --\tabularnewline
CI & 95.8\% & 95\% & -- & --\tabularnewline
\hline 
\end{tabular}

\end{table}

Table \ref{table1} shows the absolute bias of the estimated ATE $\hat{\tau}$,
true variances $v$, estimated variances $\hat{v}$, MSE, and 95\%
Wald confidence intervals. In both settings, combining EC and RE data improves accuracy and efficiency. 
% {\bf wrong! The BPP method discounts ECs by the estimated probability of trial inclusion $\mathbb{P}(S=1\mid X)$. However, it requires the full likelihood function to be correctly specified.} {\bf how the likelihood function is specified?}
The BPP procedure encounters two major challenges that may lead to increased bias. First, it relies on the full likelihood function to be accurately defined, which is an unrealistic expectation. Second, even $\mathbb{P}(S=1\mid X)$ is utilized as a power to lower the influence of the likelihood of each EC subject, $\mathbb{P}(S=1\mid X)$ itself cannot capture all the difference between ECs and CCs, leading to an increased bias. 
The MBA procedure in \citet{stuart2008matching} measures the difference between the ECs and CCs in two stages:  first they used ECs to match CCs, balancing covariates between the ECs and CCs in this process. They then determined the bias value, $\delta$, between ECs and CCs using matched groups of CCs and ECs. This $\delta$ is constant for all 
$X$ and may not be accurate. In contrast, our methods use a bias function, $b_0(X)$, which adapts to different $X$ values and accounts for all differences, irrespective of whether they arise from covariates or the outcome. Consequently, our approach is more effective than the two-stage method proposed by \citet{stuart2008matching}. As shown in Table \ref{table1}, \citet{stuart2008matching} approach has a larger bias compared to our method and the Bayesian Power Prior method due to the less accurate bias term, $\delta$.

\section{REAL DATA ANALYSIS}\label{sec:Real-Data-Analysis}

%We apply the proposed DPIE to the data from the National Supported Work (NSW) study. This study aims at evaluating the effect of a job training program on future earnings, containing an experimental sample from a randomized evaluation of the NSW program, and a nonexperimental sample from the Current Population Survey (CPS) program. $15992$ control units are included in the original CPS dataset, whereas $260$ control units are included in the NSW dataset. We use the matching procedure \citep{abadie2006large} to pair each NSW controls with $2$ CPS controls without replacement. Therefore, the RE contains $260$ random treatment units and $260$ random control units (CCs), supplemented by $520$ external control units (ECs).
We apply the proposed DPIE to data from the National Supported Work (NSW) study, which evaluates the impact of a job training program on future earnings. The dataset comprises an experimental sample for randomized evaluation and a nonexperimental sample from the Current Population Survey (CPS) program. The original CPS dataset includes $15,992$ control units. To create external controls, we employ the matching procedure \citep{abadie2006large}, matching each NSW control with $2$ CPS controls without replacement. As a result, the RE consists of $260$ random treatment units and $260$ random control units (CCs), supplemented by $520$ external control units (ECs).

This analysis includes the eight original covariates from the NSW and CPS datasets (age, education, Black, Hispanic, married, having no college degree (denoted as ``nodeg''), real earnings in 1974 (denoted as ``re74''), and real earnings in 1975 (denoted as ``re75'') as well as their 2-way interactions. The outcome of interest is the real earnings in 1978 (denoted as ``re78''). For a better regression, we divide all the real earnings (re74, re75, re78) by 1000, scale all covariates between 0 and 1, and omit variables with the same value across observations. There are 86 covariates in total, with 43 covariates in the bias function and 43 covariates in the outcome mean function. Accordingly, we use the mean of real earnings in 1978 in the REs as
the true value, $1.794$.

Table \ref{tab:real data} shows the estimated control mean $\hat{\tau}$ (reported as ``Est''), along with the standard error (reported as `` se'') and 95\% Wald confidence intervals using the proposed DPIE estimator, the SPIE estimator and the SCAD estimator only based on the RE data. The number of variables selected in the outcome mean model (reported as ``\#v\_$\bar{\mu}_{0,1}$'') and in the bias function are also reported (reported as ``\#v\_$b_{0}$''). Even though the bias function is as complex as the outcome mean model, the DPIE improves efficiency by increasing the sample size, resulting in the standard error of DPIE being smaller than that of RE in this real data scenario. On the other hand, the SPIE estimator has a larger bias than the DPIE, because the magnitude of the outcome mean function is much smaller than that of the bias function, leading to a biased estimate compared with the DPIE, which is consistent with our simulation results shown in Figure \ref{large}. Based on the DPIE estimator, the estimated average treatment effect is $1.857$.

\begin{table}[ht]
\center{}\caption{\label{tab:real data}The first panel shows estimated ${\tau}$
and corresponding standard error, bias, 95\% Wald confidence interval
and the number of selected variables in the outcome mean model. The
last two columns show the number of selected variables (\#v) in the outcome mean model $\bar\mu_{0,1}(X)$ and the bias function $b_{0}(X)$ based on the DPIE.}
\setlength{\tabcolsep}{0.5mm}{ %
\begin{tabular}{cccccc}
\hline 
 & Est  & se  & bias & \#v\_$\bar{\mu}_{0,1}$ & \#v\_$b_{0}$\tabularnewline
\hline 
DPIE  & 1.857 (0.746 , 2.969)  & 0.567 & 0.063 & 4 & 4\tabularnewline
SPIE  & 1.626 (0.582 , 2.671)  & 0.533  & 0.168 & 5 & 1\tabularnewline
RE  & 1.698 (0.455, 2.941)  & 0.634  & 0.097 & 4 & /\tabularnewline
\hline 
\end{tabular}} 
\end{table}

\section{DISCUSSION}\label{sec:Discussion}

We introduce a bias function to measure the discrepancy between the ECs and the working model in REs and use sieve estimation and feature selection techniques to handle the high-dimensional nature of the basis functions and to prevent irrelevant covariates from being included in the outcome mean model. We propose a double penalty integration estimator (DPIE) that takes advantage of the different levels of smoothness of the outcome mean and bias functions. Our results demonstrate that the DPIE is consistent, has the oracle property, and is asymptotically normal when the penalty parameters are selected appropriately. Moreover, our estimator is robust to model misspecification and is more efficient than the REs alone.

We provide a general framework with a broad class of choices for
combining multiple datasets and employing flexible penalized regression
procedures. Combining several treatments for a more accurate estimation
of the value functions in policy evaluation and individual treatment
regimes is a direct extension of our method. In addition, our outcome
$Y$ can be extended to multiple types, including survival \citep{https://doi.org/10.48550/arxiv.2201.06595}
and zero-inflation outcomes \citep{yu2021multiplicative}. In lieu
of better estimating the outcome mean function to enhance the ATE
estimate, one may directly combine the bias function and the heterogeneous
treatment effects (HTEs; \citet{yang2022elastic}), which are the causal effects of a treatment
given the characteristics of the subjects, to obtain a more accurate
estimate of the HTEs. Evaluating the HTEs is the primary question
in many domains, including precision medicine and tailored policy
recommendations \citep{https://doi.org/10.48550/arxiv.2011.08047, chu2022targeted}.
Finally, we exclusively consider the SCAD penalty in our theoretical
study. The SCAD penalty addresses consistency, oracle property, and asymptotic normality of some local minimizer of the penalized loss. However, it doesn't ensure the uniqueness of the solution or provide methods for identifying the specific local minimizer with the desired properties among a large pool of potential local minimizes \citep{zhang2010nearly}. This gap between theory and practice presents an interesting avenue for future research. To address this concern, we propose several potential approaches: \citet{fan2014strong} introduced a general procedure based on the LLA algorithm and derive a lower bound on the probability that a specific local solution exactly matches the oracle estimator, which could be applicable in real-world scenarios; \citet{kim2012global} provided conditions for determining the uniqueness of a local minimizer. Additionally, we recommend varying initial values in R's \texttt{ncvfit} function, and selecting the estimate that minimizes error. Alternatively, using unpenalized estimated covariates as initial values can be considered.
Moreover, 
%the SCAD penalty encounters the computational difficulty of the non-convex optimization problem \citep{hesterberg2008least}.
A general theoretical framework for multiple penalties, such as the adaptive Lasso \citep{zou2006adaptive} and minimax concave penalty \citep{zhang2010nearly} of double penalty selection, would therefore be desirable.

% References

\bibliography{cheng_477}
\end{document}

\onecolumn %% Turn this off if single column is desired for the supplement
\maketitle
%\hbadness=10000 \tolerance=10000 \hyphenation{en-vi-ron-ment
%in-ven-tory e-num-er-ate char-ac-ter-is-tic}
    %\renewcommand{\theequation}{\thesubsection.\arabic{equation}}
    %Wayne wants simple eq numbers so they don't have to be redone
    %use renewcommand for appendix equation numbering (B.1) etc.

%\usepackage{hangpar}
\newcommand{\lbl}[1]{\label{#1}{\ensuremath{^{\fbox{\tiny\upshape#1}}}}}
% remove % from next line for final copy
\renewcommand{\lbl}[1]{\label{#1}}
\newtheorem{lemma}{Lemma}\newtheorem{theorem}{Theorem}\newtheorem{assumption}{Assumption}\newtheorem{remark}{Remark}\newtheorem{corollary}{Corollary}\newtheorem{example}{Example}\newtheorem{definition}{Definition}[section]
\newtheorem{proof}{Proof}\newtheorem{condition}{Condition}\newtheorem{step}{Step}\newcommand{\bx}{\mathbf{x}}
\newcommand{\bz}{\mathbf{z}}
\newcommand{\bR}{\mathbf{R}}
\newcommand{\bw}{\mathbf{w}}
\newcommand{\mhat}{\hat{\mu}}                     %mu hat
\newcommand{\bmhat}{\mbox{\boldmath$\hat{\mu}$}}  %bold mu hat
\newcommand{\bs}{\mbox{\boldmath$\sigma$}}        %bold sigma
\newcommand{\bS}{\mbox{\boldmath$\Sigma$}}        %bold Sigma
\newcommand{\ch}{{\mathcal{F}}}
\newcommand{\be}{\begin{equation}}
\newcommand{\en}{\end{equation}}
\newcommand{\bea}{\begin{eqnarray}}
\newcommand{\ena}{\end{eqnarray}}
\newcommand{\ba}{\begin{array}}
\newcommand{\percent}{\%}
\newcommand{\ea}{\end{array}}
\newcommand{\dis}{\displaystyle}
\newcommand{\Pf}{\vspace{0.3 cm}\no\underline{\it Proof}\hspace{0.7 cm}}
\newcommand{\vs}{\vspace{0.6 cm}}
\newcommand{\hs}{\hspace{1 cm}}
\newcommand{\adrsq}{(1-\frac{1}{r})S_r^2}
\newcommand{\bym}{\bar{y}_m^*}
\newcommand{\byr}{\bar{y}_r}
\newcommand{\sumh}{\sum_{c=1}^C}
\newcommand{\hyi}{\hat{Y}_I}
\newcommand{\A}{A}
\newcommand{\rr}{A_R\,}
\newcommand{\hnv}{\hat{N}_v}
\newcommand{\hrv}{\hat{R}_v}
\newcommand{\byn}{\bar{y}_n}
\newcommand{\bxm}{\bar{x}_m^*}
\newcommand{\byi}{\hat{\mu}_I}
\newcommand{\bxr}{\bar{x}_r}
\newcommand{\bxn}{\bar{x}_n}
\newcommand{\bxi}{\bar{x}_I}
\newcommand{\byrv}{\bar{y}_r^v}
\newcommand{\byii}{\hat{\mu}_{I(-i)}}
\newcommand{\sumi}{\sum_{i=1}^{n}}
\newcommand{\vnaive}{V_{JK}^I}
\newcommand{\no}{\noindent}
\newcommand{\R}{{\mathcal{R}}}
\newcommand{\pop}{{\mathbf{Y}}}
\newcommand{\pr}{\mathbb{P} }
\newcommand{\indep}{\perp \!\!\! \perp}
\def\mH{\mathcal{H}}

\newcommand{\T}{\mathrm{\scriptscriptstyle T}}
\newcommand{\mis}{ {\mathrm{mis}}} 
\newcommand{\ATT}{ {\mathrm{ATT}}} 
\newcommand{\adj}{ {\mathrm{adj}}} 
\newcommand{\mat}{ {\mathrm{mat}}} 
\newcommand{\obs}{ {\mathrm{obs}}} 
\newcommand{\var}{ {\mathrm{var}}} 
\newcommand{\cov}{ {\mathrm{cov}}} 
\newcommand{\dsm}{ {\mathrm{dsm}}} 
\newcommand{\psm}{ {\mathrm{psm}}} 
\newcommand{\prog}{ {\mathrm{prog}}} 
\newcommand{\J}{ {\mathcal{J}}} 
\newcommand{\mP}{ {\mathbb{P}}} 
\newcommand{\plim}{ {\mathrm{plim}}} 
\newcommand{\F}{ {\mathcal{F}}} 
\newcommand{\rep}{ {\mathrm{rep}}} 
\newcommand{\reg}{ {\mathrm{REG}}} 
\newcommand{\nni}{ {\mathrm{NNI}}} 
\newcommand{\nnri}{ {\mathrm{NNRI}}}
\newcommand{\HT}{ {\mathrm{HT}}} 
\newcommand{\N}{ {\mathcal{N}}} 
\newcommand{\I}{ {\tau}} 
\newcommand{\It}{ \mathcal{I}} 
\newcommand{\logit}{ {\mathrm{logit}}} 
\newcommand{\de}{ {\mathrm{d}}} 
\newcommand{\mx}{ {\mathrm{m.x}}} 
\newcommand{\dm}{ {d_{V}}} 
\newcommand{\E}{ {\mathbb{E}} } 
\newcommand{\bP}{ {\mathbb{P}}} 
\newcommand{\V}{ {\mathbb{V}}} 
\newcommand{\bone}{\mathbf{1}}
\newcommand{\cU}{ {\mathcal{U}}} 
\newcommand{\sgn}{ {\mathrm{sgn}}}

\global\long\def\theequation{S\arabic{equation}}
\setcounter{equation}{0}
\global\long\def\thefigure{S\arabic{figure}}
\setcounter{figure}{0}
\global\long\def\thesection{S\arabic{section}}
\setcounter{section}{0}
\global\long\def\thetheorem{S\arabic{theorem}}
\setcounter{theorem}{0}
\global\long\def\thecondition{S\arabic{condition}}
\setcounter{condition}{0}
\global\long\def\theremark{S\arabic{remark}}
\setcounter{remark}{0}
\global\long\def\thestep{S\arabic{step}}
\setcounter{step}{0}
\global\long\def\theassumption{S\arabic{assumption}}
\setcounter{assumption}{0}

\renewcommand{\labelenumi}{\alph{enumi})}

% If your paper is accepted, change the options for the package
% aistats2022 as follows:
%
%\usepackage[accepted]{aistats2022}
%
% This option will print headings for the title of your paper and
% headings for the authors names, plus a copyright note at the end of
% the first column of the first page.

% If you set papersize explicitly, activate the following three lines:
%\special{papersize = 8.5in, 11in}
%\setlength{\pdfpageheight}{11in}
%\setlength{\pdfpagewidth}{8.5in}

% If you use natbib package, activate the following three lines:
%\usepackage[round]{natbib}
%\renewcommand{\bibname}{References}
%\renewcommand{\bibsection}{\subsubsection*{\bibname}}

% If you use BibTeX in apalike style, activate the following line:
%\bibliographystyle{apalike}

The supplementary material is structured as follows:
Section \ref{subsec:Regularity-conditions} includes regularity assumptions.
Section \ref{subsec:Proof} provides proofs for the main theorems.
Section \ref{sec:Toy-example} presents the toy example mentioned in Section 2.
Section \ref{plots} displays additional figures for the first simulation study in this section. The simulation codes are in the \url{https://github.com/yuwen997/simulation-codes}.

\section{Regularity conditions}\label{subsec:Regularity-conditions}

In this section, we provide the same assumptions on likelihood functions and density functions as those in
\citet{fan2004nonconcave,white1982maximum}. 
To utilize their languages, we specify a working density function $f(Y,p,\theta)=f(Y,p,\beta,\delta)$\footnote{We will use $\theta$ and $(\beta,\delta)$ exchangeably for clarity in different contexts.} such that 
the minimizer $\theta_*=(\beta_{*},\delta_{*})$ of $\E\{Y-h(A,X,S;\beta,\delta)\}^{2}$, where $h(A,X,S;\beta,\delta)=\beta_{{\rm int}}+\beta_{A}A+\beta_{X}^{{\rm T}}p_{\mu}(X)+(1-S)\delta^{{\rm T}}p_{b}(X)$,
is also the minimizer of the Kullback-Leibler Information Criterion, $KLIC(g:f,\beta,\delta)=\mathbb{E}\left[\log\left\{ g(Y,p)/f(Y,p,\beta,\delta)\right\} \right]$. 
Here, $g(Y,p)$ is the true density function of $(Y,p)$ and does not depend on the unknown parameter $(\beta,\delta)$.
A straightforward choice of $f(Y,p,\beta,\delta)$ is the Gaussian density function with mean $h(A,X,S;\beta,\delta)$. 
It is also important to mention that Theorem 2 and Theorem 3 can be extended to incorporate various working density functions $f$, given that $f$ and $g$ satisfy assumptions \ref{f1}--\ref{f3}. 
%Subsequently, there's a local maximizer $\hat\theta$ that allows the penalized quasi-log-likelihood function $Q(\theta)$ to converge to $\tilde\theta_*$, where $\tilde\theta_*$ minimizes the KLIC between $f$ and $g$.

For a better understanding of these conditions, an informal summary of assumptions S1--S9 is provided here. Assumptions S1--S7 align with A1--A7 in \citet{white1982maximum}, ensuring MLE estimator consistency and asymptotic normality in both misspecified and correct models. S8--S9 resemble F--G in \citet{fan2004nonconcave}, bounding $f$ moments.
\begin{assumption}\label{f1}The
independent random vectors $(p_{i},Y_{i}),$ $i=1,\ldots,N,$ have
common joint distribution function $G$ on $\Upsilon$, a measurable
Euclidean space, with measurable Radon-Nikodym density $g=dG/d\mu.$ 

\end{assumption}

\begin{assumption}\label{f1-1}The family of distribution functions
$F(Y_{1},p,\theta)$ has Radon-Nikodym densities $f(y,p,\theta)=dF(y,p,\theta)/d\mu$
which are measurable in $(y,p)$ for every $\theta$ in $\Theta,$
a compact subset of Euclidean space, and continuous in $\theta$
for every $(y,p)$ in $\Upsilon.$

\end{assumption}

\begin{assumption}\label{f1-2} (a) $\E\{\log g(Y,p)\}$ exists and $\vert\log f(y,p,\theta)\vert\leq m(y,p)$
for all $\theta$ in $\Theta$, where $m$ is integrable with respect
to $G$; (b) $KLIC(g:f,\theta)$ has a unique minimum at $\theta_{*}$
in $\Theta$.

\end{assumption}

\begin{assumption}\label{f1-3} $\partial\log f(y,p,\theta)/\partial\theta_{j},j=1,\ldots,K,$
are measurable functions of $(y,p)$ for each $\theta$ in $\Theta$
and continuously differentiable functions of $\theta$ for each $(y,p)$
in $\Upsilon.$ \end{assumption}

\begin{assumption}\label{f1-4} $\vert\partial^{2}\log f(y,p,\theta)/\partial\theta_{i}\partial\theta_{j}\vert$
and $\vert\partial\log f(y,p,\theta)/\partial\theta_{i}\cdot\partial\log f(y,p,\theta)/\partial\theta_{j}\vert$,
$i,j=1,\ldots,K$ are dominated by functions integrable with respect
to $G$ for each $\theta$ in $\Theta$ and $(y,p)$ in $\Upsilon.$
\end{assumption}

\begin{assumption}\label{f1-5} Define matrix 
\begin{align*}
A\left(\theta\right) & =-{\mathbb{E}}\left\{\frac{\partial^{2}\log f\left(Y_{1},p_{1},\theta\right)}{\partial\theta_{j}\partial\theta_{k}}\right\}>0,\\
B(\theta) & =\E\left[\left\{ \frac{\partial\log f\left(Y_{1},p_{1},\theta\right)}{\partial\theta}\right\} \left\{ \frac{\partial\log f\left(Y_{1},p_{1},\theta\right)}{\partial\theta}\right\} ^{{\rm T}}\right],
\end{align*}
and (a) $\theta_{*}$ is interior to $\Theta$; (b) $B(\theta_{*})$
is nonsingular; (c) $\theta_{*}$ is a regular point of $A(\theta).$

\end{assumption}

\begin{assumption}\label{f1-6} $\vert\partial\left\{ \text{\ensuremath{\partial f(y,p,\theta)/\partial\theta_{i}\cdot f(y,p,\theta)}}\right\} /\partial\theta_{j}\vert$,
$i,j=1,\ldots,K$ are dominated by functions integrable with respect
to $\mu$ for all $\theta$ in $\Theta$ and the minimal support of
$f(y,p,\theta)$ does not depend on $\theta.$ \end{assumption}

\begin{comment}
For every $N$ the observations $\left(p_{1},Y_{1}\right),\ldots,\left(p_{N},Y_{N}\right)$
are i.i.d. with the probability density $f\left(Y_{1},p_{1},\theta\right)$,
which has a common support, and the model is identifiable. Furthermore,
the first derivatives of the likelihood function satisfies the equation
\[
\E_{\theta}\left\{ \frac{\partial\log f\left(Y_{1},p_{1},\theta\right)}{\partial\theta_{j}}\right\} =0\ {\rm for}\ j=1,...,K.
\]
\end{comment}

\begin{assumption}\label{f2}Define matrix 
\begin{align*}
C(\theta) & =A(\theta)B^{-1}(\theta)A(\theta).
\end{align*}

Assume matrix $A(\theta)$ and $B(\theta)$ satisfy the following conditions 
\begin{align*}
0 & <C_{1}<\lambda_{\min}\left\{ A\left(\theta\right)\right\} \leq\lambda_{\max}\left\{ A\left(\theta\right)\right\} <C_{2}<\infty\ \ {\rm for}\ {\rm all}\ N,\\
0 & <C_{1}^{*}<\lambda_{\min}\left\{ B\left(\theta\right)\right\} \leq\lambda_{\max}\left\{ B\left(\theta\right)\right\} <C_{2}^{*}<\infty\ \ {\rm for}\ {\rm all}\ N,
\end{align*}

and for $j,k=1,\ldots,K,$ 
\[
\E_{\theta}\left\{ \frac{\partial\log f\left(Y_{1},p_{1},\theta\right)}{\partial\theta_{j}}\frac{\partial\log f\left(Y_{1},p_{1},\theta\right)}{\partial\theta_{k}}\right\} ^{2}<C_{3}<\infty
\]

and 
\[
\E_{\theta}\left\{ \frac{\partial^{2}\log f\left(Y_{1},p_{1},\theta\right)}{\partial\theta_{j}\partial\theta_{k}}\right\} ^{2}<C_{4}<\infty.
\]

\end{assumption}

\begin{assumption}\label{f3}
There is a large enough open subset $\omega_{N}$ of $\Theta\in R^{K}$
which contains the parameter point $\theta_{*}$, such that for almost
all $\left(p_{i},Y_{i}\right)$ the density admits all third derivatives
$\partial f(Y_{i},p_{i},\theta)/\partial\theta_{j}\theta_{k}\theta_{l}$
for all $\theta\in\omega_{N}$. Furthermore, there are functions $M_{jkl}$
such that 
\[
\left\vert\frac{\partial\log f\left(Y_{i},p_{i},\theta\right)}{\partial\theta_{j}\partial\theta_{k}\partial\theta_{l}}\right\vert\leq M_{jkl}\left(Y_{i},p_{i}\right)
\]
for all $\theta\in\omega_{N},$ and 
\[
\E_{\theta}\left\{ M_{jkl}^{2}\left(Y_{i},p_{i}\right)\right\} <C_{5}<\infty
\]
for all $K,N,j,k$ and $l$.
\end{assumption}

%\begin{assumption}\label{f4}
%The working density function $f(Y,p,\beta,\delta)$ satisfies that the minimizers $(\beta_{*},\delta_{*})$ of $\E\{Y-h(A,X,S;\beta,\delta)\}^{2}$
%are also the minimizers  of $KLIC(g:f,\beta,\delta)=\mathbb{E}\left[\log\left\{ g(Y,p)/f(Y,p,\beta,\delta)\right\} \right]$, where $g(Y,p)$ is the true density function and does not depend on the unknown parameter $(\beta,\delta)$. 
%\end{assumption}

\section{Proof}\label{subsec:Proof}

In this section, we provide proof of Theorems 1--4.
Define $L\left(\theta\right)=L(\beta,\delta)=\sum_{i=1}^{N} \ln f\left(Y_{i},p_{i,}\beta,\delta\right) .$
Subsequently, the penalized quasi-likelihood function is $Q(\theta)=L(\theta)-N\sum_{i=1}^{K_{1}}P_{\lambda_{1,i}}\left(\vert\beta\vert\right)-N\sum_{i=1}^{K_{2}}P_{\lambda_{2,i}}\left(\vert\delta\vert\right).$
% Assume $\beta_{*}=\begin{pmatrix}\beta_{*1}\\
% \beta_{*2}
% \end{pmatrix},\ \delta_{*}=\begin{pmatrix}\delta_{*1}\\
% \delta_{*2}
% \end{pmatrix}$ where $\beta_{*1}\neq0$ with $s_{1}$ dimensions, $\delta_{*1}\neq0$
% with $s_{2}$ dimensions, $\beta_{*2}=0$ with $K_{1}-s_{1}$ dimensions
% and $\delta_{*2}=0$ with $K_{2}-s_{2}$ dimensions. Further, let
% $\theta_{*}=\begin{pmatrix}\theta_{*1}\\
% \theta_{*2}
% \end{pmatrix}$ , where $\theta_{*1}=\begin{pmatrix}\beta_{*1}\\
% \delta_{*1}
% \end{pmatrix}\neq0$ with $s=s_{1}+s_{2}$ dimensions and $\theta_{*2}=\begin{pmatrix}\beta_{*2}\\
% \delta_{*2}
% \end{pmatrix}=0$ with $K-s$ dimensions.

\subsection{Theorem S1}

Theorem S1 was previously demonstrated in \citet{lorentz1966approximation}
and \citet{chen2007large}, and we restate it here.

\begin{theorem}\label{theorem1}For any unknown function $f:$$\mathbb{R}^{d}\rightarrow\mathbb{R}$,
assuming function $f\left(\cdot\right)$ is $t$ times continuously
differentiable. Let $K=(q+1)^{d}$ where $x_{1},\ldots,x_{d}$ are
at least up to power $q$, and let $r^{K}(x)$ be the $K$-dimension
power series basis function, $R\left(x\right)=A_{K}r^{K}\left(x\right)$
where $A_{K}$ is the matrix such that $\E\left\{ R\left(X\right)R^{{\rm T}}\left(X\right)\right\} =\mathcal{I}$
where $\mathcal{I}$ is the identity matrix. Then there is a $K$-vector
$\theta$ such that on the compact set $\mathcal{X}$, $\sup_{x\in\mathcal{X}}\vert f\left(x\right) -R^{{\rm T}}\left(x\right)\theta\vert=O\left(K^{-t/d}\right)$.
\end{theorem}

\subsection{Proof for Theorem 1}

Under the combined dataset, the ANCOVA working model for $\E(Y\mid A,X,S)$  is
\begin{align*}
\bar{\mu}_{A,S}(X;\beta) & =\beta_{{\rm int}}+\beta_{A}A+\beta_{X}^{{\rm T}}p_{\mu}(X)+(1-S)b_{0}(X)=\bar{\mu}_{A,1}(X;\beta) +(1-S)b_{0}(X).
\end{align*}

%We can rewrite the model for $\E(\tilde{Y}\mid A,X,S)$, where
Denote $\tilde{Y}=Y-(1-S)b_{0}(X)$. %, as
%$$
%  \beta_{{\rm int}}+\beta_{A}A+\beta_{X}^{{\rm T}}p_{\mu}(X)
%  =\beta^{{\rm T}}p_{\mu}
%  =\bar{\mu}_{A,1}(X;\beta).
%$$
%Note, $\bar{\mu}_{A,1}(X;\beta)=\beta^{{\rm T}}p_{\mu}$.
%Recall $\bar{\mu}_{A,1}(X)=\beta_{0}+\beta_{A}A+\beta_{X}^{{\rm T}}p_{\mu}(X).$
%Following the same proof of the theorem for Linear Models in \citet{rosenblum2009using},
%the ordinary least squares estimate of $\beta$ is asymptotically
%normal and converges in probability to 
%Thus, $\beta_{*}$ minimizes 
Thus, the least squares loss function is
$\E\{{Y}-\bar{\mu}_{A,S}(X;\beta)\}^{2}=\E\{\tilde{Y}-\bar{\mu}_{A,1}(X;\beta)\}^{2}$, which can be further evaluated as
\begin{align*}
\E\{\tilde{Y}-\bar{\mu}_{A,1}(X;\beta)\}^{2} & =\E\left\{\tilde{Y}-\E\left(\tilde{Y}\mid A,X,S\right)+\E\left(\tilde{Y}\mid A,X,S\right)-\bar{\mu}_{A,1}(X;\beta)\right\}^{2}\\
 & =\E\left\{\tilde{Y}-\E\left(\tilde{Y}\mid A,X,S\right)\right\}^{2}+\E\left\{ \E\left(\tilde{Y}\mid A,X,S\right)-\bar{\mu}_{A,1}(X;\beta)\right\} ^{2}\\
 & =\E\left\{\tilde{Y}-\E\left(\tilde{Y}\mid A,X,S\right)\right\}^{2}+\E\left[\E\left\{ \E\left(\tilde{Y}\mid A,X,S\right)-\bar{\mu}_{A,1}(X;\beta)\right\} ^{2}\mid S\right]\\
 & =\E\left\{\tilde{Y}-\E\left(\tilde{Y}\mid A,X,S\right)\right\}^{2}+\E \left[\left\{ \E\left(\tilde{Y}\mid A,X,S=1\right)-\bar{\mu}_{A,1}(X;\beta)\right\} ^{2}\mid S=1 \right] \mathbb{P}(S=1)\\
 & +\E\left[ \left\{ \E\left(\tilde{Y}\mid A=0,X,S=0\right)-\bar{\mu}_{0,1}(X;\beta)\right\} ^{2} \mid S=0 \right]\mathbb{P}(S=0).
\end{align*}

We now show that $\beta_*=\left(\beta_{\rm{int*}},\beta_{A*},\beta_{X*}^{{\T}}\right)^\T$ as the minimizer of $\E [\{Y-\bar{\mu}_{A,1}(X;\beta)\}^{2} \mid S=1]$ also minimizes the above loss function. 
In particular, we show that $\beta_*$ minimizes both
\begin{equation}\label{T1}
\E \left[\left\{ \E\left(\tilde{Y}\mid A,X,S=1\right)-\bar{\mu}_{A,1}(X;\beta)\right\} ^{2}\mid S=1 \right], \end{equation}
and
\begin{equation} \label{T2}\E\left[ \left\{ \E\left(\tilde{Y}\mid A=0,X,S=0\right)-\bar{\mu}_{0,1}(X;\beta)\right\} ^{2} \mid S=0 \right].\end{equation}

First, by the definition, $\beta_{*}$ minimizes 
\begin{eqnarray*}
\E[\{Y-\bar{\mu}_{A,1}(X;\beta)\}^{2}\mid S=1] & = & \E[\{Y-\E(Y\mid A,X,S=1)+\E(Y\mid A,X,S=1)-\bar{\mu}_{A,1}(X;\beta)\}^{2}\mid S=1]\\
 & = & \E[\{Y-\E(Y\mid A,X,S=1)\}^{2}\mid S=1]\\
 &  & +\E[\{\E(Y\mid A,X,S=1)-\bar{\mu}_{A,1}(X;\beta)\}^{2}\mid S=1].
\end{eqnarray*}
Thus, $\beta_{*}$ minimizes $\E[\{\E(Y\mid A,X,S=1)-\bar{\mu}_{A,1}(X;\beta)\}^{2}\mid S=1]$, which equals \eqref{T1}. 

Second, by the definition of $b_{0}(X),$ we have 
\begin{align*}
\E\left(\tilde{Y}\mid A=0,X,S=0\right)-\bar{\mu}_{0,1}(X;\beta) & =\E\left\{ Y-b_{0}(X)\mid A=0,X,S=0\right\} -\bar{\mu}_{0,1}(X;\beta)\\
 & =\E\left\{ Y-\E\left(Y\mid A=0,X,S=0\right)+\bar{\mu}_{0,1}(X;\beta_*)\mid A=0,X,S=0\right\} -\bar{\mu}_{0,1}(X;\beta)\\
 & =\E\left(Y\mid A=0,X,S=0\right)-\E\left(Y\mid A=0,X,S=0\right)+\bar{\mu}_{0,1}(X;\beta_*)-\bar{\mu}_{0,1}(X;\beta)\\
 & =\bar{\mu}_{0,1}(X;\beta_*)-\bar{\mu}_{0,1}(X;\beta).
\end{align*}
Thus, $\beta_*$ minimizes \eqref{T2}. The proof for the first part of Theorem 1 is now complete.

Now, we show that $\beta_{A*}$ identifies $\tau$. By the definition, $\beta_*$ minimizes
$$
  \E\left[\left\{ \E\left(Y\mid A,X,S=1\right)-\bar{\mu}_{A,1}(X;\beta)\right\} ^{2}\mid S=1\right]\\
	=\E\left[\left\{ \mu_{A,1}(X)-\bar{\mu}_{A,1}(X;\beta)\right\} ^{2}\mid S=1\right].
$$

% On the other hand, $\beta_{*}$ minimizes 
% \begin{align*}
%     \E[\{Y-\bar{\mu}_{A,1}(X;\beta)\}^{2}\mid S=1]	&=\E[\{Y-\mu_{A,1}(X)+\mu_{A,1}(X)-\bar{\mu}_{A,1}(X;\beta)\}^{2}\mid S=1]\\
% 	&=\E[\{Y-\mu_{A,1}(X)\}^{2}\mid S=1]+\E[\{\mu_{A,1}(X)-\bar{\mu}_{A,1}(X;\beta)\}^{2}\mid S=1]\\
% 	&=\E[\{Y-\mu_{A,1}(X)\}^{2}\mid S=1]+\E\{\mu_{A,1}(X)-\bar{\mu}_{A,1}(X;\beta)\}^{2}.
% \end{align*} Therefore $\tilde{\beta}_{*}=\beta_{*}$,which both minimizes $\E\left\{ \mu_{A,1}(X)-\bar{\mu}_{A,1}(X;\beta)\right\} ^{2}$. Similar in  \citep{wang2021model}, 

Thus, the first derivative of
$\E\left[\left\{ \mu_{A,1}(X)-\bar{\mu}_{A,1}(X;\beta)\right\} ^{2} \mid S=1\right]$ evaluated at $\beta_*$ is a vector of zeros. Because $\bar{\mu}_{A,1}(X;\beta_*)$ includes an intercept term,  $\beta_{*}$
satisfies $\E\left\{ \mu_{A,1}(X)-\bar{\mu}_{A,1}(X;\beta_*)\mid S=1\right \} =0.$ 
Based on this result, we have $\tau=\E\left\{ \mu_{1,1}(X)-\mu_{0,1}(X)\mid S=1 \right\}  =\E\left\{ \bar{\mu}_{1,1}(X)-\bar{\mu}_{0,1}(X)\mid S=1 \right\}  =\beta_{A*}.$
%Similarly, in the REs, we have $\tau=\beta_{A*}.$
The proof for the second part of Theorem 1 is complete.

\subsection{Proof for Theorem 2}

Assume $g(Y,p)$ is the true density function, $f(Y,p,\theta)$ is
the working density function.
By the choice of $f$, $\theta_{*}=(\beta_*^{\rm{T}},\delta_*^{\rm{T}})^{\rm{T}}$ minimizing  $\E\{Y-h(A,X,S;\beta,\delta)\}^{2}$ also minimizes 
\[
KLIC(g:f,\theta)=\mathbb{E}\left[\log\left\{ g(Y,p)/f(Y,p,\theta)\right\} \right].
\]

We follow the similar proofs in \citet{fan2004nonconcave}, let $a_{N}=\sqrt{K}\left(N^{-1/2}+\alpha_{N}\right)$
and set $\|\mathbf{u}\|_{2}=C$, where $C$ is a large enough constant,
our aim is to show that for any given $\epsilon$ there is a large
constant $C$ such that, for large $N$ we have 
\[
\mathbb{P}\left\{ \sup_{\|\mathbf{u}\|_{2}=C}Q(\theta_{*}+a_{N}\mathbf{u})<Q(\theta_{*})\right\} \geq1-\epsilon.
\]

This implies that with probability tending to $1$ there is a local
maximum $\hat{\theta}$ in the call $\left\{ \theta_{*}+a_{N}\mathbf{u}:\|\mathbf{u}\|_{2}\leq C\right\} $
such that $\|\hat{\theta}-\theta_{*}\|_{2}=O_{p}\left(a_{N}\right).$
Because $P_{\lambda_{1}}(0)=P_{\lambda_{2}}(0)=0,$ We have 
\begin{align*}
D(\mathbf{u}) & =Q(\theta_{*}+a_{N}\mathbf{u})-Q(\theta_{*})\\
 & \leq\underbrace{L(\theta_{*}+a_{N}\mathbf{u})-L(\theta_{*})}_{\left(I\right)}\\
 & \underbrace{-N\sum_{j=1}^{s_{1}}\left\{ P_{\lambda_{1}}\left(\vert\beta_{*,j}+a_{N}u_{1j}\vert\right)-P_{\lambda_{1}}\left(\vert\vert\beta_{*,j}\vert\right)\right\} }_{\left(II\right)}\\
 & \underbrace{-N\sum_{j=1}^{s_{2}}\left\{ P_{\lambda_{2}}\left(\vert\delta_{*,j}+a_{N}u_{2j}\vert\right)-P_{\lambda_{2}}\left(\vert\delta_{*,j}\vert\right)\right\} }_{\left(III\right)}\\
 & :=\left(I\right)+\left(II\right)+\left(III\right),
\end{align*}
where $\mathbf{u}^{{\rm T}}=(\mathbf{u_{1}^{{\rm T}},\mathbf{u}_{2}^{{\rm T}})}$
with $\text{\ensuremath{\mathbf{u_{1}}}}$ as $K_{1}$ dimensions
and $\text{\ensuremath{\mathbf{u_{2}}}}$ as $K_{2}$ dimensions.
First for $(II)$ we have 
\begin{align*}
(II) & =-\sum_{j=1}^{s_{1}}\left[Na_{N}P_{\lambda_{1}}^{'}\left(\vert\beta_{*,j}\vert\right)\sgn(\beta_{*,j})u_{1j}+Na_{N}^{2}P_{\lambda_{1}}^{''}\left(\beta_{*,j}\right)u_{1j}^{2}\left\{ 1+o(1)\right\} \right]\\
 & :=I_{1}+I_{2},\\
\vert I_{1}\vert & \leq\sum_{j=1}^{s_{1}}\vert Na_{N}P_{\lambda_{1}}^{'}\left(\vert\beta_{*,j}\vert\right)\sgn(\beta_{*,j})u_{1j}\vert\leq\sqrt{s_{1}}Na_{N}\alpha_{N}\|\mathbf{u}_{1}\|_{2}\leq Na_{N}^{2}\|\mathbf{u}\|_{2},\\
\vert I_{2}\vert & =\sum_{j=1}^{s_{1}}Na_{N}^{2}P_{\lambda_{1}}^{''}(\vert\beta_{*,j}\vert)u_{1j}^{2}\left\{ 1+o(1)\right\} \leq2\max_{1\leq j\leq s_{1}}P_{\lambda_{1}}^{''}(\vert\beta_{*,j}\vert)Na_{N}^{2}\|\mathbf{u}\|_{2}^{2}.
\end{align*}
Similarly for $(III).$ Then for $(I)$ we have 
\begin{align*}
(I) & =a_{N}\text{\ensuremath{\nabla^{{\rm T}}L(\theta_{*})\mathbf{u}+\frac{1}{2}\mathbf{u}^{{\rm T}}\nabla^{2}L(\theta_{*})\mathbf{u}a_{N}^{2}+\frac{1}{6}\nabla^{{\rm T}}\left\{ \mathbf{u}^{{\rm T}}\nabla^{2}L(\theta^{*})\mathbf{u}\right\} \mathbf{u}a_{N}^{3}}}\\
 & :=I_{3}+I_{4}+I_{5},
\end{align*}
with the same proof in Theorem 1 in \citet{fan2004nonconcave}, by
condition \ref{f2}, we have 
\begin{align*}
\vert I_{3}\vert & =\vert a_{N}\nabla^{{\rm T}}L(\theta_{*})\mathbf{u}\vert\leq a_{N}\|\nabla^{{\rm T}}L(\theta_{*})\|_{2}\|\mathbf{u}\|_{2}=O_{p}(a_{N}^{2}N)\|\mathbf{u}\|_{2}.\\
I_{4} & =\frac{1}{2}\mathbf{u}^{{\rm T}}\left\{ \frac{1}{N}\left(\left[\nabla^{2}L(\theta_{*})-\E\left\{ \nabla^{2}L(\theta_{*})\right\} \right]\right)\right\} \mathbf{u}Na_{N}^{2}\\
 & -\frac{1}{2}\mathbf{u}^{{\rm T}}A(\theta_{*})\mathbf{u}Na_{N}^{2}\\
 & =-\frac{Na_{N}^{2}}{2}\mathbf{u}^{{\rm T}}A(\theta_{*})\mathbf{u}+o_{p}(1)Na_{N}^{2}\|\mathbf{u}\|_{2}^{2}.
\end{align*}
By condition \ref{f3} and $K^{4}/N\rightarrow0$ and $K^{2}\alpha_{N}\rightarrow0$
as $N\rightarrow\infty$, we have 
\begin{align*}
\vert I_{5}\vert & =\vert\frac{1}{6}\sum_{i,j,k=1}^{K}\frac{\partial L(\theta^{*})}{\partial\theta_{i}\partial\theta_{j}\partial\theta_{k}}u_{i}u_{j}u_{k}a_{N}^{3}\vert\\
 & \leq\frac{1}{6}\sum_{l=1}^{N}\left\{ \sum_{i,j,k=1}^{K}M_{ijk}^{2}(Y_{i},p_{i})\right\} ^{1/2}\|\mathbf{u}\|_{2}^{3}a_{N}^{3}\\
 & =o_{p}(Na_{N}^{2})\|\mathbf{u}\|_{2}^{2}.
\end{align*}
Therefore, by Assumption 5 and allowing $\|\mathbf{u}\|_{2}$ to be
large enough, all $I_{1},I_{2},I_{3},I_{5}$ and $(III)$ are dominated
by $I_{4}$, which is negative, therefore proves $\|\hat{\theta}-\theta_{*}\|_{2}=O_{p}\left\{ \sqrt{K}\left(N^{-1/2}+a_{N}\right)\right\} .$
Further we have $\max\left\{ \|\hat{\beta}-\beta_{*}\|_{2},\|\hat{\delta}-\delta_{*}\|_{2}\right\} \leq\|\hat{\theta}-\theta_{*}\|_{2}=O_{p}\left\{ \sqrt{K}\left(N^{-1/2}+a_{N}\right)\right\} .$
For the SCAD penalty, it is clear that $a_{N}=O_{p}\left(N^{-1/2}\right)$,
therefore $\hat{\beta}$ and $\hat{\delta}$ are root-$\left(N/K\right)$-consistent
penalized likelihood estimators exist with probability tending to
1, and no requirements are imposed on the convergence rate of $\lambda_{1}$
and $\lambda_{2}$.

\subsection{Proof for Theorem 3}

We follow the similar proofs in \citet{fan2004nonconcave}. we first
show that the nonconcave penalized estimator possesses the sparsity
property $\hat{\theta}_{2}=0$ by the following lemma.

\begin{lemma}\label{lemma1}Assume Assumption 5, Assumption \ref{f1}--\ref{f3}
are satisfied, if $\lambda_{1},\lambda_{2}\rightarrow0,$ $\sqrt{N/K}\lambda_{1}\rightarrow\infty,$
$\sqrt{N/K}\lambda_{2}\rightarrow\infty$, and $K^{5}/N\rightarrow0$
as $N\rightarrow\infty,$ then first show that with probability tending
to $1$, for any given $\theta_{1}$ satisfying $\|\theta_{1}-\theta_{*1}\|_{2}=O_{p}\left(\sqrt{K/N}\right)$
and any constant $C,$ 
\[
Q\left\{ (\theta_{1}^{{\rm T}},0)^{{\rm T}}\right\} =\max_{\|\theta_{2}\|_{2}\leq C(K/N)^{1/2}}Q\left\{ (\theta_{1}^{{\rm T}},\theta_{2}^{{\rm T}})^{{\rm T}}\right\} .
\]

\end{lemma}

Proof: Let $\epsilon=C\sqrt{K/N}.$ It is sufficient to show that
with probability tending to $1$ as $N\rightarrow\infty,$ for any
$\theta_{1}-\theta_{*1}=O_{p}\left(\sqrt{K/N}\right)$ we have for
$j=s+1,\ldots,K,$

\[
\begin{array}{ccc}
\frac{\partial Q(\theta)}{\partial\theta_{j}} & <0 & {\rm for}\ 0<\theta_{j}<\epsilon,\\
\frac{\partial Q(\theta)}{\partial\theta_{j}} & >0 & {\rm for}\ -\epsilon<\theta_{j}<0.
\end{array}
\]

By Taylor expansion,

\begin{align*}
\frac{\partial Q(\theta)}{\partial\theta_{j}} & =\frac{\partial L(\theta)}{\partial\theta_{j}}-NP_{\lambda}^{'}(\vert\theta_{j}\vert)\sgn\left(\theta_{j}\right)\\
 & =\frac{\partial L(\theta_{*})}{\partial\theta_{j}}+\sum_{l=1}^{K}\frac{\partial^{2}L(\theta_{*})}{\partial\theta_{j}\partial\theta_{l}}(\theta_{l}-\theta_{*,l})\\
 & +\sum_{l,k=1}^{K}\frac{\partial^{3}L(\theta^{*})}{\partial\theta_{j}\partial\theta_{l}\partial\theta_{k}}(\theta_{l}-\theta_{*,l})\left(\theta_{k}-\theta_{*,k}\right)\\
 & -NP_{\lambda}^{'}(\vert\theta_{j}\vert)\sgn\left(\theta_{j}\right)\\
 & :=I_{1}+I_{2}+I_{3}+I_{4},
\end{align*}
where $\theta^{*}$ lies between $\theta$ and $\theta_{*}$, and
$P_{\lambda}^{'}(\vert\theta_{j}\vert)=P_{\lambda_{1}}^{'}(\vert\beta_{j}\vert)$
for $j=s+1,\ldots,K_{1}-s_{1}+s,$ and $P_{\lambda}^{'}(\vert\theta_{j}\vert)=P_{\lambda_{2}}^{'}(\vert\delta_{j}\vert)$
for $j=K_{1}-s_{1}+s+1,\ldots,K.$

Following the same proof in \citet{fan2004nonconcave}, we prove $I_{1}+I_{2}+I_{3}=O_{p}\left(\sqrt{NK}\right).$
First, $I_{1}=O_{p}\left(\sqrt{N}\right)=O_{p}\left(\sqrt{NK}\right).$
Also, 
\begin{align*}
I_{2} & =\sum_{l=1}^{K}\left(\frac{\partial^{2}L(\theta_{*})}{\partial\theta_{j}\partial\theta_{l}}-\E\left\{ \frac{\partial^{2}L(\theta_{*})}{\partial\theta_{j}\partial\theta_{l}}\right\} \right)\left(\theta_{l}-\theta_{*,l}\right)\\
 & +\text{\ensuremath{\sum_{l=1}^{K}\E\left\{ \frac{\partial^{2}L(\theta_{*})}{\partial\theta_{j}\partial\theta_{l}}\right\} }}\left(\theta_{l}-\theta_{*,l}\right)\\
 & :=S_{1}+S_{2}.
\end{align*}

Using the Cauchy-Schwarz inequality and $\|\theta-\theta_{*}\|_{2}=O_{p}\left(K/N\right),$ we
have 
\begin{align*}
\vert S_{2}\vert & =\vert N\sum_{l=1}^{K}A(\theta_{*})(j,l)(\theta_{l}-\theta_{*,l})\vert\\
 & \leq NO_{p}\left(\sqrt{\frac{K}{N}}\right)\left\{ \sum_{l=1}^{K}A^{2}(\theta_{*})(j,l)\right\} ^{1/2}.
\end{align*}
By Assumption \ref{f2}, as the eigenvalues of the $A(\theta)$ are
bounded, we have $S_{2}=O_{p}\left(\sqrt{NK}\right).$ On the other
hand, 
\[
\vert S_{1}\vert\leq\|\theta-\theta_{*}\|_{2}\left(\sum_{l=1}^{K}\left[\frac{\partial^{2}L(\theta_{*})}{\partial\theta_{j}\partial\theta_{l}}-\E\left\{ \frac{\partial^{2}L(\theta_{*})}{\partial\theta_{j}\partial\theta_{l}}\right\} \right]^{2}\right)^{1/2}.
\]
By Assumption \ref{f2}, we have 
\[
\left(\sum_{l=1}^{K}\left[\frac{\partial^{2}L(\theta_{*})}{\partial\theta_{j}\partial\theta_{l}}-\E\left\{ \frac{\partial^{2}L(\theta_{*})}{\partial\theta_{j}\partial\theta_{l}}\right\} \right]^{2}\right)^{1/2}=O_{p}\left(\sqrt{NK}\right).
\]
Therefore $S_{1}=O_{p}\left(\sqrt{NK}\right)$ and $I_{2}=O_{p}\left(\sqrt{NK}\right)$.
Further, 
\begin{align*}
I_{3} & =\sum_{l,k=1}^{K}\left[\frac{\partial^{3}L(\theta^{*})}{\partial\theta_{j}\partial\theta_{l}\partial\theta_{k}}-\E\left\{ \frac{\partial^{3}L(\theta^{*})}{\partial\theta_{j}\partial\theta_{l}\partial\theta_{k}}\right\} \right](\theta_{l}-\theta_{*,l})\left(\theta_{k}-\theta_{*,k}\right)\\
 & +\sum_{l,k=1}^{K}\E\left\{ \frac{\partial^{3}L(\theta^{*})}{\partial\theta_{j}\partial\theta_{l}\partial\theta_{k}}\right\} (\theta_{l}-\theta_{*,l})\left(\theta_{k}-\theta_{*,k}\right)\\
 & :=S_{3}+S_{4}.
\end{align*}

By Assumption \ref{f3}, $\vert S_{4}\vert\leq C_{5}^{1/2}NK\|\theta-\theta_{*}\|_{2}^{2}=O_{p}(K^{2})=o_{p}\left(\sqrt{NK}\right).$
Further, 
\[
S_{3}^{2}\leq\sum_{l,k=1}^{K}\left[\frac{\partial^{3}L(\theta^{*})}{\partial\theta_{j}\partial\theta_{l}\partial\theta_{k}}-\E\left\{ \frac{\partial^{3}L(\theta^{*})}{\partial\theta_{j}\partial\theta_{l}\partial\theta_{k}}\right\} \right]^{2}\|\theta-\theta_{*}\|_{2}^{4},
\]
where under the Assumption \ref{f3} and Assumption 5, $S_{3}=O_{p}\left\{ \left(NK^{2}\frac{K^{2}}{N^{2}}\right)^{1/2}\right\} =o_{p}\left(\sqrt{NK}\right).$
Then 
\[
I_{1}+I_{2}+I_{3}=O_{p}\left(\sqrt{NK}\right).
\]

Because we focus on the SCAD penalty, \citet{fan2004nonconcave} illustrates
that under Assumption 5, the SCAD penalty satisfies that 
\begin{align*}
\lim\inf_{N\rightarrow+\infty}\lim\inf_{\beta\rightarrow0+}P_{\lambda_{1}}^{'}(\beta)/\lambda_{1} & >0\\
\lim\inf_{N\rightarrow+\infty}\lim\inf_{\delta\rightarrow0+}P_{\lambda_{2}}^{'}(\delta)/\lambda_{2} & >0,
\end{align*}
therefore from 
\[
\frac{\partial Q(\theta)}{\partial\theta_{j}}=N\lambda\left\{ -\frac{P_{\lambda}^{'}\left(\vert\theta_{j}\vert\right)}{\lambda}\sgn({\theta_{j}})+O_{p}\left(\sqrt{\frac{K}{N}}/\lambda\right)\right\} ,
\]
where $\lambda=\lambda_{1}$ if $j=s+1,\ldots,K_{1}-s_{1}+s$ and
$\lambda=\lambda_{2}$ if $j=K_{1}-s_{1}+s+1,\ldots,K,$ and $P_{\lambda}^{'}(\vert\theta_{j}\vert)=P_{\lambda_{1}}^{'}(\vert\beta_{j}\vert)$
for $j=s+1,\ldots,K_{1}-s_{1}+s,$ and $P_{\lambda}^{'}(\vert\theta_{j}\vert)=P_{\lambda_{2}}^{'}(\vert\delta_{j}\vert)$
for $j=K_{1}-s_{1}+s+1,\ldots,K,$ the sign of $\theta_{j}$ completely
determines the sign of $\partial Q(\theta)/\partial\theta_{j}.$ We
complete the proof of Lemma \ref{lemma1}.

By Lemma \ref{lemma1} we prove $\hat{\theta}_{2}=\begin{pmatrix}\hat{\beta}_{2}\\
\hat{\delta}_{2}
\end{pmatrix}=0$. Then we prove the part 2.

Let 
\begin{align*}
\Sigma & ={\rm diag}\left\{ P_{\lambda}^{''}\left(\theta_{*,1}\right),\ldots,P_{\lambda}^{''}\left(\theta_{*,s}\right)\right\} \\
 & ={\rm diag}\left\{ P_{\lambda_{1}}^{''}\left(\beta_{*,1}\right),\ldots,P_{\lambda_{1}}^{''}\left(\beta_{*,s_{1}}\right),P_{\lambda_{2}}^{''}\left(\delta_{*,1}\right),\ldots,\ P_{\lambda_{2}}^{''}\left(\delta_{*,s_{2}}\right)\right\} 
\end{align*}
and 
\begin{align*}
b & =\left\{ P_{\lambda}^{'}\left(\vert\theta_{*,1}\vert\right)\sgn\left(\theta_{*,1}\right),\ldots,P_{\lambda}^{'}\left(\vert\theta_{*,s}\vert\right)\sgn\left(\theta_{*,s}\right)\right\} ^{{\rm T}}\\
 & =\left\{ P_{\lambda_{1}}^{'}\left(\vert\beta_{*,1}\vert\right)\sgn\left(\beta_{*,1}\right),\ldots,P_{\lambda_{1}}^{'}\left(\vert\beta_{*,s_{1}}\vert\right)\sgn\left(\beta_{*,s_{1}}\right),\ P_{\lambda_{2}}^{'}\left(\vert\delta_{*,1}\vert\right)\sgn\left(\delta_{*,1}\right),\ldots,P_{\lambda_{2}}^{'}\left(\vert\delta_{*,s_{2}}\vert\right)\sgn\left(\delta_{*,s_{2}}\right)\right\} ^{{\rm T}}.
\end{align*}

If we can show that 
\[
\left\{ A(\theta_{*1})+\Sigma\right\} \left(\hat{\theta}_{1}-\theta_{*1}\right)+b=\frac{1}{N}\nabla L(\theta_{*1})+o_{p}(N^{-1/2}),
\]
then 
\begin{align*}
 & \sqrt{N}WA^{-1/2}(\theta_{*1})\left\{ A(\theta_{*1})+\Sigma\right\} \left[\hat{\theta}_{1}-\theta_{*1}+\left\{ A(\theta_{*1})+\Sigma\right\} ^{-1}b\right]\\
= & \frac{1}{\sqrt{N}}WA^{-1/2}(\theta_{*1})\nabla L(\theta_{*1})+o_{p}\left\{ WA^{-1/2}(\theta_{*1})\right\} \\
= & \frac{1}{\sqrt{N}}WA^{-1/2}(\theta_{*1})\nabla L(\theta_{*1})+o_{p}(1).
\end{align*}

Let $R_{i}=\frac{1}{\sqrt{N}}WA^{-1/2}(\theta_{*1})\nabla L_{i}(\theta_{*1}),$
$i=1,\ldots,N.$ Following the same proof in \citet{fan2004nonconcave},
for any $\epsilon,$ we have 
\begin{align*}
\sum_{i=1}^{N}\E\|R_{i}\|_{2}^{2}\bone\left\{ \|R_{i}\|_{2}>\epsilon\right\}  & =N\mathbb{\mathbb{E}}\|R_{1}\|_{2}^{2}\bone\left\{ \|R_{1}\|_{2}>\epsilon\right\} ,\\
 & \leq N(\mathbb{\mathbb{E}}\|R_{1}\|_{2}^{4})^{1/2}\left\{ \mathbb{P}\left(\|R_{1}\|_{2}>\epsilon\right)\right\} ^{1/2}.
\end{align*}

By Assumption \ref{f2} and $WW^{{\rm T}}\rightarrow G,$ we obtain
\[
\mathbb{P}\left(\|R_{1}\|_{2}>\epsilon\right)\leq\frac{\mathbb{E}\|WA^{-1/2}(\theta_{*1})\nabla L_{1}(\theta_{*1})\|_{2}^{2}}{N\epsilon^{2}}=O(N^{-1})
\]
 and 
\begin{align*}
\mathbb{\mathbb{E}}\|R_{1}\|_{2}^{4} & =\frac{1}{N^{2}}\mathbb{E}\|WA^{-1/2}(\theta_{*1})\nabla L_{1}(\theta_{*1})\|_{2}^{4}\\
 & \leq\frac{1}{N^{2}}\lambda_{{\rm max}}(WW^{{\rm T}})\lambda_{{\rm max}}\left\{ A^{-1}(\theta_{*1})\right\} \mathbb{E}\|\nabla^{{\rm T}}L_{1}(\theta_{*1})\nabla L_{1}(\theta_{*1})\|_{2}^{2}\\
 & \leq O\left(\frac{K^{2}}{N^{2}}\right).
\end{align*}

Thus, we have 
\[
\sum_{i=1}^{N}\E\|R_{i}\|_{2}^{2}\bone\left\{ \|R_{i}\|_{2}>\epsilon\right\} =O\left(N\frac{K}{N}\frac{1}{\sqrt{N}}\right)=o(1).
\]

and

\[
\sum_{i=1}^{N}\cov(R_{i})=\cov\left\{ WA^{-1/2}(\theta_{*1})\nabla L_{1}(\theta_{*1})\right\} =WA^{-1/2}(\theta_{*1})B(\theta_{*1})A^{-1/2}(\theta_{*1})W^{{\rm T}},
\]
so that the $R_{i}$ satisfies the conditions of the Lindeberg-Feller
central limit theorem. Further, using the Taylor expansion on $\nabla Q(\hat{\theta}_{1})$
at the point $\theta_{*1}$, we have 
\begin{align*}
 & \frac{1}{N}\left[\left\{ \nabla^{2}L(\theta_{*1})-\nabla^{2}P_{\lambda}(\theta_{1}^{**})\right\} \left(\hat{\theta}_{1}-\theta_{*1}\right)-\nabla P_{\lambda}\left(\theta_{*1}\right)\right]\\
= & -\frac{1}{N}\left[\nabla L(\theta_{*1})+\frac{1}{2}\left(\hat{\theta}_{1}-\theta_{*1}\right)^{{\rm T}}\nabla^{2}\left\{ \nabla L\left(\theta_{1}^{*}\right)\left(\hat{\theta}_{1}-\theta_{*1}\right)\right\} \right],
\end{align*}
where $\theta_{1}^{*}$ and $\theta_{1}^{**}$ lie between $\hat{\theta}_{1}$
and $\theta_{*1}.$ Now define 
\[
\mathcal{L}:=\nabla^{2}L(\theta_{*1})-\nabla^{2}P_{\lambda}(\theta_{1}^{**})
\]
and 
\[
C:=\frac{1}{2}\left(\hat{\theta}_{1}-\theta_{*1}\right)^{{\rm T}}\nabla^{2}\left\{ \nabla L\left(\theta_{1}^{*}\right)\left(\hat{\theta}_{1}-\theta_{*1}\right)\right\} .
\]
Following the proof in \citet{fan2004nonconcave}, under Assumption
\ref{f3} and Assumption 5 and by the Cauchy--Schwarz inequality,
we have $\|1/N\mathcal{C}\|_{2}^{2}=o_{p}\left(1/N\right).$  Further, 
we have 
\[
\lambda_{i}\left\{ \frac{1}{N}\mathcal{L}+A(\theta_{*1})+\Sigma\right\} =o_{p}\left(\frac{1}{\sqrt{K}}\right),i=1,\ldots,s,
\]
where $\lambda_{i}(M)$ is the $i$th eigenvalue of a symmetric matrix
$M$. Therefore, 
\[
\left\{ \frac{1}{N}\mathcal{L}+A(\theta_{*1})+\Sigma\right\} \left(\hat{\theta}_{1}-\theta_{*1}\right)=o_{p}\left(\frac{1}{\sqrt{N}}\right).
\]
Then, we have $\left\{ A(\theta_{*1})+\Sigma\right\} \left(\hat{\theta}_{1}-\theta_{*1}\right)+b=\frac{1}{N}\nabla L(\theta_{*1})+o_{p}(N^{-1/2}),$
and finally we have 
\begin{align*}
\sqrt{N}WA^{-1/2}(\theta_{*1})\left\{ A(\theta_{*1})+\Sigma\right\} \left[\hat{\theta}_{1}-\theta_{*1}+\left\{ A(\theta_{*1})+\Sigma\right\} ^{-1}b\right]\\
\rightarrow\mathcal{N}(0,WA^{-1/2}(\theta_{*1})\left[B(\theta_{*1})-\mathbb{E}\left\{ \nabla L_{1}(\theta_{*1})\right\} \right]A^{-1/2}(\theta_{*1})W^{{\rm T}}\text{.}
\end{align*}
Further, based on the SCAD penalty, $\Sigma=0$ and $b=0$, therefore,
we have 
\[
\sqrt{N}WA^{1/2}\left(\theta_{*1}\right)\left(\hat{\theta}_{1}-\theta_{*1}\right)\rightarrow\mathcal{N}\left(0,WA^{-1/2}(\theta_{*1})B(\theta_{*1})A^{-1/2}(\theta_{*1})W^{{\rm T}}\right).
\]
 If the model is correctly specified, i.e., $g(Y,p_{i})=f(Y,p_{i},\theta)$
for some $\theta\in\Theta$, then $\theta_{0}=\theta_{*}$, and 
\[
\sqrt{N}WI^{1/2}\left(\theta_{01}\right)\left(\hat{\theta}_{1}-\theta_{01}\right)\rightarrow\mathcal{N}\left(0,WW^{{\rm T}}\right).
\]

We finish the second part. 

%\subsection{Comments on Theorem 2 and Theorem 3}

%It's important to clarify that if our focus isn't primarily on the least square loss, we can adjust the assumptions on $f$ and $g$ from \ref{f1}--\ref{f3} to only satisfy Assumptions \ref{f1}--\ref{f3} for Theorem 2 and Theorem 3. The critical point is that for any other $f$ and $g$ fulfilling Assumptions \ref{f1}--\ref{f3}, a local maximizer $\hat{\theta}$ of the $Q(\theta)$ exists such that $\hat{\theta}$ converges to $\theta_*$, where $\theta_*$ minimizes the KLIC between $f$ and $g$, and also exhibits the oracle property and asymptotic normality. 

%However, in the main paper, our primary concern is the least square loss, and we further constrain $f$ to meet assumption \ref{f4} in addition to assumptions \ref{f1}--\ref{f3}. Assumption \ref{f4} implies that the parameters minimizing the KLIC also minimize a specific form: the least square form. This is because in the main paper, we are particularly interested in ANCOVA least square estimates.

\subsection{Proof for Theorem 4}

The consistency and the asymptotic normality of $\hat\tau$ follow from Theorems 2 and 3. Here we only prove the calculation of the variance. 
If $\V(\epsilon)=\sigma^{2},$
then the influence function of $\theta$ is 
\[
\phi_{\theta}(X,Y,S)=\E\left\{ \begin{pmatrix}p_{\mu}\\
(1-S)p_{b}^{{\rm }}(X)
\end{pmatrix}\begin{pmatrix}p_{\mu}^{{\rm T}}, & (1-S)p_{b}^{{\rm T}}(X)\end{pmatrix}\right\} ^{-1}\begin{pmatrix}p_{\mu}\\
(1-S)p_{b}^{{\rm }}(X)
\end{pmatrix}(Y-p^{{\rm T}}\theta).
\]
% Then the asymptotic variance of $\hat{\beta}$ is
% \begin{align*}
% \V(\hat{\beta}) & =\E\left\{ \phi_{\theta}(X,Y,S)\phi_{\theta}^{{\rm T}}(X,Y,S)\right\} _{11}\\
%  & =\sigma^{2}\E\left\{ \begin{pmatrix}p_{\mu}^{{\rm }}p_{\mu}^{{\rm T}} & (1-S)p_{\mu}^{{\rm }}p_{b}^{{\rm T}}(X)\\
% (1-S)p_{b}(X)p_{\mu}^{{\rm T}} & (1-S)p_{b}^{{\rm }}(X)p_{b}(X)^{{\rm T}}
% \end{pmatrix}\right\} _{11}^{-1}\\
%  & =\sigma^{2}\E\left[Sp_{\mu}^{{\rm }}p_{\mu}^{{\rm T}}+(1-S)p_{\mu}^{{\rm }}p_{\mu}^{{\rm T}}-(1-S)\left\{ p_{\mu}^{{\rm }}p_{b}^{{\rm T}}(X)\right\} \left\{ p_{b}^{{\rm }}(X)p_{b}^{{\rm T}}(X)\right\} ^{-1}\left\{ p_{b}(X)p_{\mu}^{{\rm T}}\right\} \right]^{-1}.
% \end{align*}

Then the asymptotic variance of $\hat{\beta}$ is
\begin{align*}
\V(\hat{\beta}) & =\E\left\{ \phi_{\theta}(X,Y,S)\phi_{\theta}^{{\rm T}}(X,Y,S)\right\} _{11}\\
 & =\sigma^{2}\E\left\{ \begin{pmatrix}p_{\mu}^{{\rm }}p_{\mu}^{{\rm T}} & (1-S)p_{\mu}^{{\rm }}p_{b}^{{\rm T}}(X)\\
(1-S)p_{b}(X)p_{\mu}^{{\rm T}} & (1-S)p_{b}^{{\rm }}(X)p_{b}(X)^{{\rm T}}
\end{pmatrix}\right\} _{11}^{-1}\\
 & =\sigma^{2}\left[\E (Sp_{\mu}^{{\rm }}p_{\mu}^{{\rm T}})+\E \{ (1-S)p_{\mu}^{{\rm }}p_{\mu}^{{\rm T}}\}-\E \left\{ (1-S) p_{\mu}^{{\rm }}p_{b}^{{\rm T}}(X)\right\} \E \left\{ (1-S) p_{b}^{{\rm }}(X)p_{b}^{{\rm T}}(X)\right\} ^{-1}\E \left\{ (1-S) p_{b}(X)p_{\mu}^{{\rm T}}\right\} \right]^{-1}.
\end{align*}

On the other hand, under the RE data, the ANCOVA working model is
\begin{align*}
Y & =\beta_{{\rm int}}+\beta_{A}A+\beta_{X}^{{\rm T}}p_{\mu}(X)+\epsilon,
\end{align*}
Similarly, the asymptotic variance of $\hat{\beta}_{{\rm RE}}$ is
\begin{align*}
\V(\hat{\beta}_{{\rm RE}}) & =\sigma^{2}\E\left(Sp_{\mu}p_{\mu}^{{\rm T}}\right)^{-1}.
\end{align*}
Then by Holder inequality, 
% \begin{align*}
%  & \E\left\{ (1-S)p_{\mu}^{{\rm }}p_{\mu}^{{\rm T}}\right\} \\
%  & -\E\left[(1-S)\left\{ p_{\mu}^{{\rm }}p_{b}^{{\rm T}}(X)\right\} \left\{ p_{b}^{{\rm }}(X)p_{b}^{{\rm T}}(X)\right\} ^{-1}\left\{ p_{b}(X)p_{\mu}^{{\rm T}}\right\} \right]\geq0,
% \end{align*}
\begin{align*}
 & \E\left\{ (1-S)p_{\mu}^{{\rm }}p_{\mu}^{{\rm T}}\right\} \\
 & -\E \left\{ (1-S) p_{\mu}^{{\rm }}p_{b}^{{\rm T}}(X)\right\} \E \left\{ (1-S) p_{b}^{{\rm }}(X)p_{b}^{{\rm T}}(X)\right\} ^{-1}\E \left\{ (1-S) p_{b}(X)p_{\mu}^{{\rm T}}\right\}\geq0,
\end{align*}
where the equality holds iff $p_{\mu}=Mp_{b}(X)$ for some matrix
$M$. However, it is worth noting that $p_\mu=\{1,A,p_\mu(X)^{\rm{T}}\}^{\rm{T}}$ includes the treatment information, while $p_b(X)$ does not. As a result, there is no such matrix $M$ that satisfies the equation $p_{\mu}=Mp_{b}(X)$. 
Therefore, we have $\V( \hat{\beta}) < \V( \hat{\beta}_{{\rm RE}}) ,$ and $\V(\hat{\tau}) < \V(\hat{\tau}_{{\rm RE}}).$
% and the equality holds iff $p_{\mu}=Mp_{b}(X)$ for some matrix
% $M$. 

\section{A toy example} \label{sec:Toy-example}

\begin{figure}[h]
\center{} \includegraphics[scale=0.15]{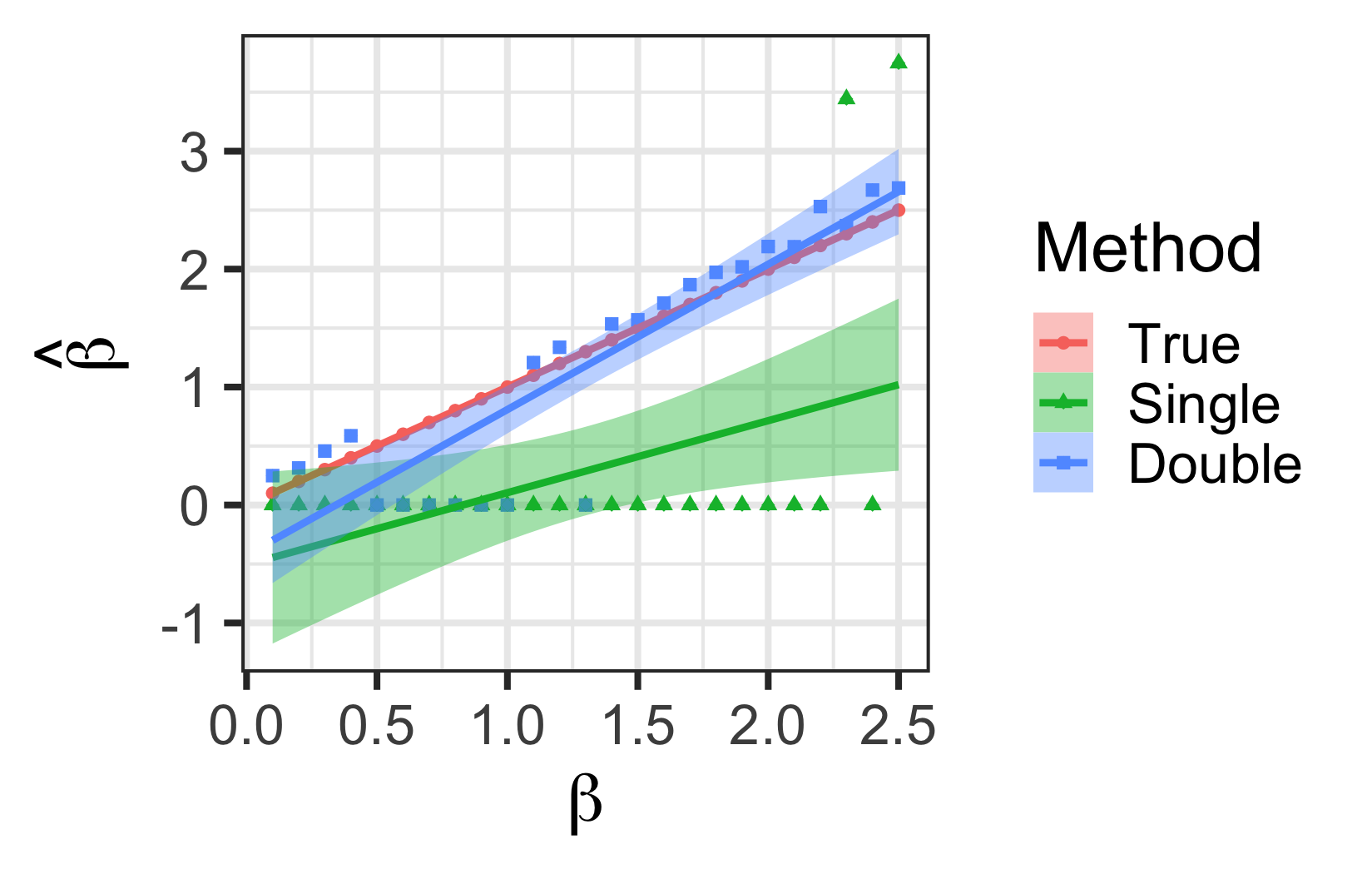}

\caption{\label{example2-1}The smoothed linear regression between $\hat{\beta}$
and $\beta$ with the 95\% confidence intervals as the shaded area
and $(\beta,\hat{\beta})$ as the points.}
\end{figure}

Consider the case $y=x^{{\rm T}}\beta+(1-s)x^{{\rm T}}\delta+\epsilon,\ x=\left(x_{1},\ldots,x_{50}\right)^{{\rm T}}\in\mathbb{R}^{50},\ \beta^{{\rm T}}=(1,\ldots,50)/10,\ \delta^{{\rm T}}=(1,\ldots,50)\times30,\epsilon\sim\mathcal{N}(0,1)$,
where $s$ denotes the zero-one indicator variable that determines
whether the observation belongs to the REs, for simplicity, we assume
$x^{{\rm T}}\beta$ is the correct outcome mean function of the REs,
and $x^{{\rm T}}\delta$ is the bias function reflecting the difference
between ECs and REs, i.e., if $\delta=0$, then the observed covariates
capture all confounders in the ECs and REs and thus the exchangeability
assumption is valid. For didactic purposes, the magnitude of $\delta$
is much larger than the magnitude of $\beta$. Using the same regularization
parameter appears to assign the same weight for $\beta$ and $\delta$,
thus any penalty regularization methods tend to omit small signals
$\beta$ and only pick up big signals $\delta.$ Therefore, in order
to make penalizations between different parameters comparable, it
is crucial to add regularizations to $\beta$ and $\delta$ separately.
Figure \ref{example2-1} shows the smoothed linear regression between
$\hat{\beta}$ and $\beta$ after applying the single-penalty regularization
(denoted as ``Single'') and the double-penalty regularizations (denoted
as ``Double'') to select variables and refitting the model using
selected variables, where double penalties make $\hat{\beta}$ more
accurate than the single penalties. 

\section{Plots}\label{plots}

\begin{figure}[h]
\center{}\includegraphics[scale=0.35]{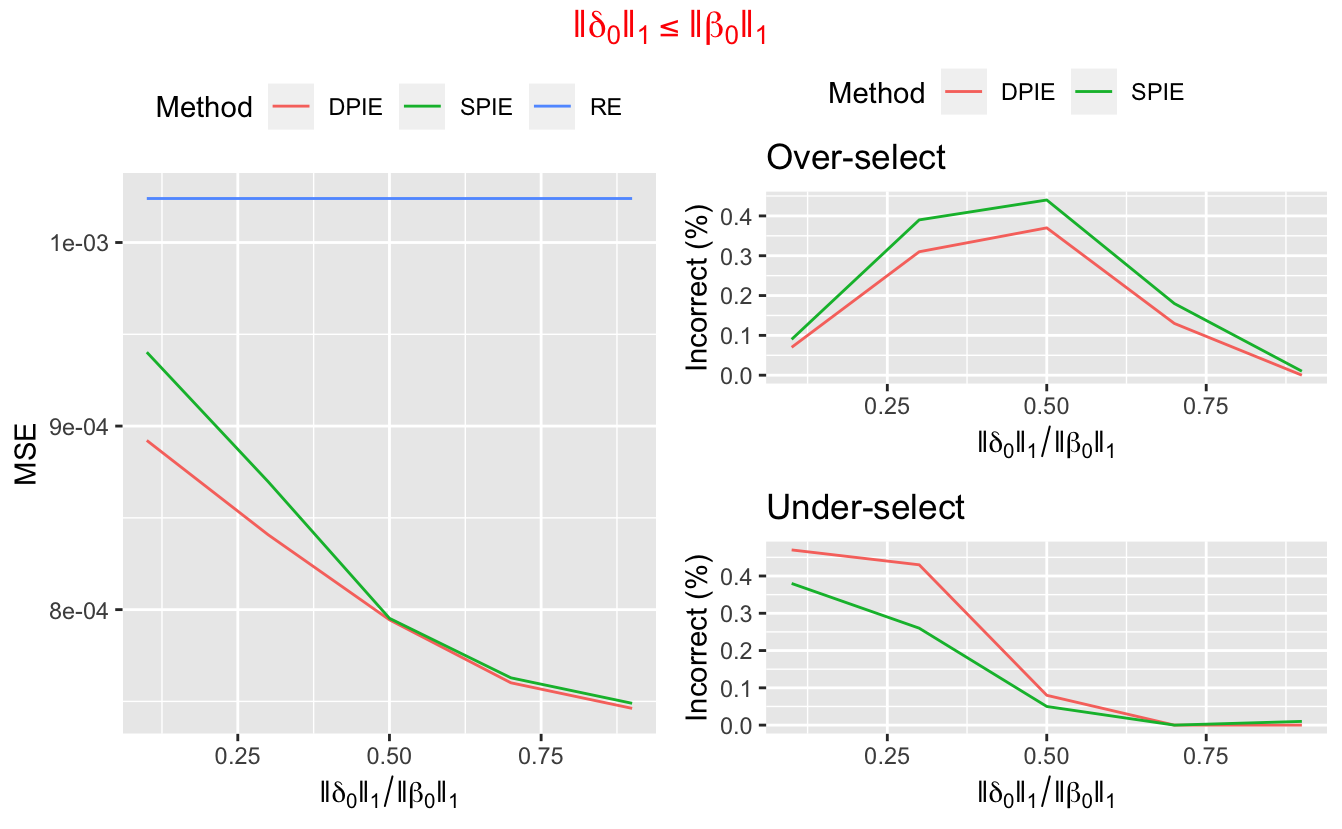}
\caption{\label{small}Simulation results based on 100 Monte Carlo times. The
left panel shows the MSE versus the magnitude ratio between $\delta$
and $\beta$. The right panel shows the percentage of wrongly choosing
more and less parameters, separately.}
\end{figure}

\begin{figure}[h]
\center{}\includegraphics[scale=0.35]{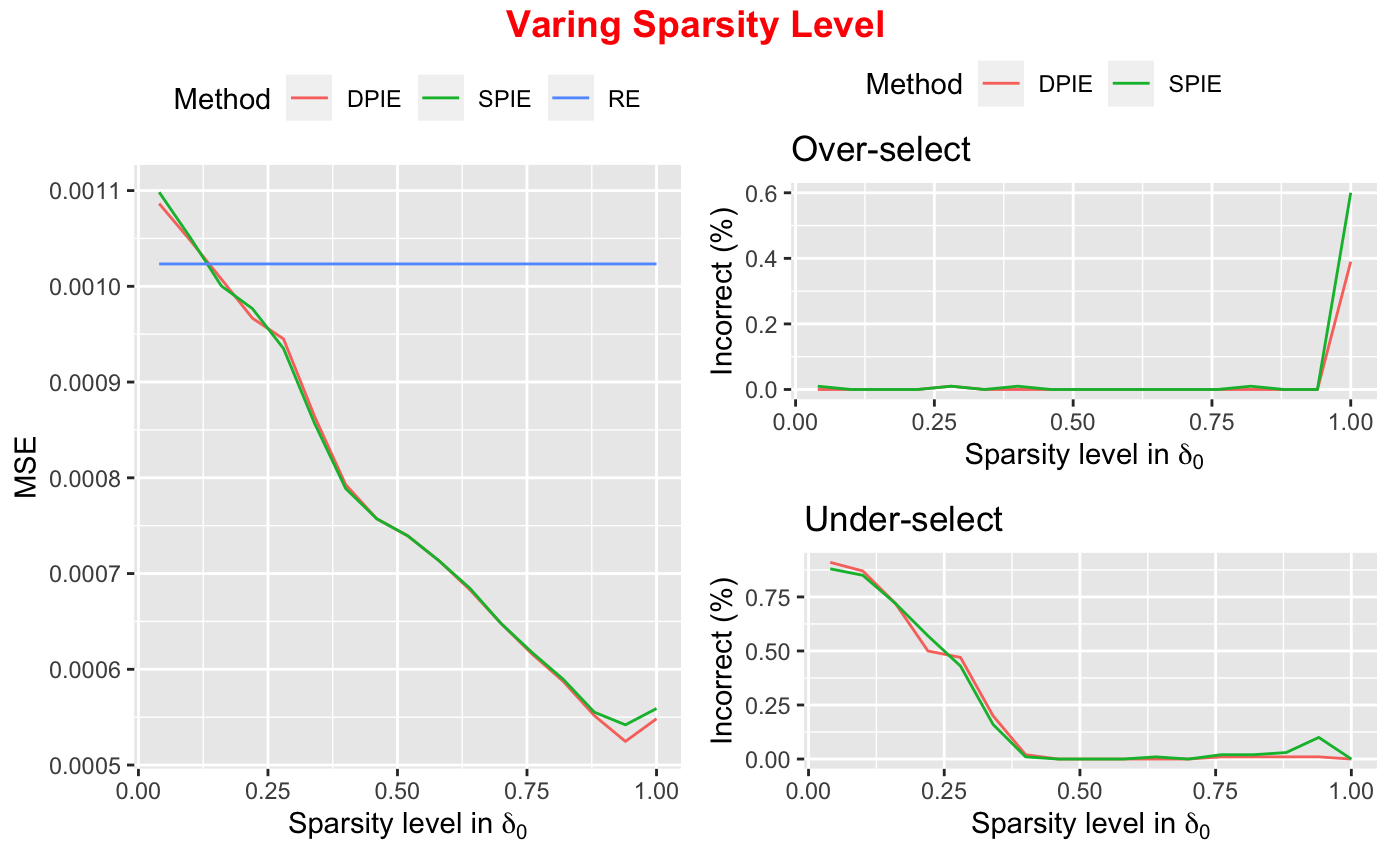}
\caption{\label{sparsity}Simulation results based on 100 Monte Carlo times.
The left panel shows the MSE versus the sparsity level in $\delta$.
The right panel shows the percentage of wrongly choosing more and
less parameters, separately.}
\end{figure}

We present the extra figures for the first simulation study in this
section. Figure \ref{small} shows the results for the case $\|\beta_{0}\|_{1}\geq \|\delta_{0}\|_{1}$
with half of the parameters in $\delta_{0}$ set to zero; $\|\beta_{0}\|_{1}=c\|\delta_{0}\|_{1},c=1,3,5,7,9;$
and Figure \ref{sparsity} shows the results for varying the sparsity
level of $\delta_{0}$ when setting $\|\beta_{0}\|_{1}=\|\delta_{0}\|_{1}$
with the x-axis as the ratio of variables in $\delta_{0}$ that are equal to zero.
Each figure shows the MSE results and the percentage of Under-select
and Over-select. Figure \ref{small} shows the SPIE has a larger MSE
compared to the DPIE, which is consistent with the theoretical results.
On the other hand, the changes in the sparsity of $\delta_{0}$ affect
little on the results. This finding also consists of the theoretical
result, where we only need to restrain the magnitude of different
parameters to guarantee the consistency and oracle properties. 

\bibliography{cheng_477-supp}